\documentclass[fleqn,aps, prb, twocolumn,showpacs,floatfix,longbibliography,nofootinbib]{revtex4-2}

\usepackage{amsmath,amssymb,amsfonts,float,graphics,epsfig,epstopdf,color,verbatim,tabularx,bm,multirow,appendix,hyperref} 
 
\usepackage{amsthm}
\theoremstyle{definition}

\newtheorem{lemma}{Lemma}

\usepackage{amsmath}
\usepackage{amssymb}
\usepackage{mathtools}
\usepackage{graphicx}
\usepackage{lipsum}
\usepackage{bm}
\usepackage{hyperref}
\usepackage{url}
\usepackage[utf8]{inputenc}
\usepackage{slashed,bbm}
\usepackage{graphics,psfrag,epsfig}
\usepackage{dsfont}
\usepackage{setspace}
\usepackage{wasysym}
\usepackage{slashed}
\usepackage{lipsum}
\usepackage{braket}
\usepackage{physics}
\usepackage{enumerate}
\usepackage{makecell}
\usepackage{xcolor}
\usepackage[normalem]{ulem}
\usepackage{tikz}
\usetikzlibrary{arrows.meta}

\definecolor{dark-red}{rgb}{0.4,0.15,0.15}
\definecolor{dark-blue}{rgb}{0.15,0.15,0.4}
\definecolor{medium-blue}{rgb}{0,0,0.5}
\definecolor{regionAC}{RGB}{121,169,245}
\definecolor{regionB}{RGB}{255,145,143}
\hypersetup{
colorlinks, linkcolor={dark-blue},
citecolor={dark-blue}, urlcolor={medium-blue}
}

\newcommand{\be}{\begin{equation}}
\newcommand{\ee}{\end{equation}}
\newcommand{\bea}{\begin{eqnarray}}
\newcommand{\eea}{\end{eqnarray}}

\renewcommand{\i}{\text{i}}

\begin{document}

\title{
Renormalization group and long-range conditional mutual information
\\
in hierarchical models
}

\author{Yu-Hsueh Chen
}
\affiliation{Department of Physics, University of California at San Diego, La Jolla, California 92093, USA\\
Kavli Institute for Theoretical Physics, University of California, Santa Barbara, California 93106, USA
}

\begin{abstract} 
A departure of a mixed quantum state from a local Gibbs description is generally invisible to local observables but can be detected by the conditional mutual
information (CMI).   
Here we study the relationship between the renormalization group (RG) and CMI, and in particular, how RG constrains CMI.
We first show that the CMI between nonadjacent regions $A$ and $C$, conditioned on the buffer region $B$, is UV-finite whenever the
state admits a locally reversible RG with a fixed on-site Hilbert space
dimension. 
We then study two hierarchical models that have long-range CMI and yet admit a
simple RG description.  
The first model has a divergent Markov length at every temperature
$0<T<\infty$ but nevertheless flows to an
infinite-temperature product state under RG. 
The second model satisfies the local Markov condition while
violating the global one and is stable against weak noise. 
At the critical noise strength, the two-point CMI 
decays only polynomially as a function of the system size, while the two-point mutual information vanishes.  
\end{abstract}

\maketitle

\section{Introduction}

A nonequilibrium process need not produce a state with a local Gibbs description.
How do we tell whether a state admits such a description? And when it does not, how do we quantify its departure from one?
A key observation points toward an answer: in \emph{any} classical Gibbs state of a local Hamiltonian, the correlations between two distant regions are mediated entirely by what lies in between, even at a critical point~\cite{clifford1971markov}.
The conditional mutual information (CMI) $I(A:C|B)=I(A:BC)-I(A:B)$, where $I(X:Y)$ denotes the mutual information between $X$ and $Y$, precisely measures the correlations between $A$
and $C$ that are not mediated by the intervening region $B$, and therefore quantifies the departure from a local Gibbs state.
The same structure holds approximately for quantum states~\cite{hayden2004structure,petz1986sufficient,fawzi2015quantum,junge2018universal,brown2012quantum,kato2019quantum,chen2025quantum}.
This makes the CMI a natural diagnostic for genuinely nonequilibrium phenomena~\cite{sang2025stability,sang2025mixed,negari2026spacetime,zhang2025conditional,yang2025topological,lloyd2025diverging,zhang2025stability,chen2025local,vijay2025information,negari2026critical,ma2025circuit,li2026unified}, including transitions that no physical observable detects~\cite{dennis2002,lee2023quantum,fan2023diagnostics,bao2023mixed}. 
In this work, we investigate the relation between the CMI and the 
renormalization group (RG), and then use the RG to study states with long-range
CMI that do not admit a local Gibbs description. 
This relation raises several questions, which we address in turn.

The essence of the RG is to turn a seemingly complicated problem into a flow in a parameter space, with fixed points capturing the universal behavior.
Interestingly, which fixed points can be connected by a flow is itself constrained
~\cite{zamolodchikov1986irreversibility,Friedan2004boundary,komargodski2011renormalization,myers2010,casini2012,patil2025shannon}.
One famous example is the $c$-theorem~\cite{zamolodchikov1986irreversibility}, which provides a $C$-function that decreases along the RG flow and equals the central charge at fixed points.
Similar ideas have since been formulated in terms of information-theoretic quantities~\cite{Casini04,casini2012,casini2016g,casini2019irreversibility,Casini2023entropic,harper2024g}. 
In fact, in the ground state of a $(1+1)$-dimensional CFT, the CMI of three adjacent intervals is proportional to the central charge, so that the $c$-theorem can be phrased directly in terms of the CMI.
Recently, Ref.~\cite{chen2026constraints} extended this idea to nonequilibrium systems, showing that the scaling function associated with CMI is an RG monotone under certain assumptions.

In a rough sense, however, the RG is simply a coarse graining of the short-distance degrees of freedom while retaining the long-distance physics. 
This perspective admits many different RG schemes, from classical block spins~\cite{kadanoff1966scaling,niemeijer1973wilson,Wilson75} to tensor network schemes for pure states~\cite{White92,vidal2007entanglement} and for mixed states~\cite{cirac2017matrix,sang2024mixed}, and which scheme is preferable is a case-by-case question.
Such freedom sits uneasily with the universal statements above. Is the $C$-function, the CMI, or any of the RG monotones above monotonically decreasing under \emph{any} microscopic RG scheme?  

The key observation is that the entropic RG monotones~\cite{Casini04,casini2012,casini2016g,casini2019irreversibility,Casini2023entropic,harper2024g}, as well as Zamolodchikov's original derivation~\cite{zamolodchikov1986irreversibility}, are usually established in two steps.
First, one shows that a quantity defined as a function of the probe
scale $R$ decreases monotonically as $R$ increases.  This step is typically
rigorous and independent of the RG scheme.  Second, one \textit{assumes}
that this quantity is \emph{invariant} under an RG transformation in which $R$
is rescaled at the same time (which in infinitesimal form is the
Callan--Symanzik equation).
This second step requires the quantity to be UV-finite.  In particular, it is what underlies
the statement in Ref.~\cite{chen2026constraints} that the CMI or its associated scaling function is an RG
monotone.
How, then, does one show that the CMI of a given state is UV-finite?

Although model dependent, a microscopic RG scheme has the merit of establishing phase equivalence constructively.
This is familiar from pure states, where finite-depth local unitary circuits, together with wave-function renormalization, define the equivalence classes~\cite{verstraete2005renormalization,vidal2007entanglement,chen2010local,chen2011classification}. 
Channels, unlike unitaries, are not generically invertible, so the mixed-state analogue is two-way: two states lie in the same phase when a pair of local channel circuits connects them, one in each direction~\cite{coser2019classification,sang2024mixed,sang2025stability}.
A more restrictive definition further requires the connecting circuit to be \emph{locally reversible}~\cite{sang2025mixed}, and this is where the CMI enters: a local Lindbladian evolution along which the Markov length, the scale on which the CMI decays, stays finite can be locally reversed~\cite{sang2025stability}.
However, we point out that a finite Markov length is sufficient but not necessary for local reversibility. 
For instance, long-range interactions can be irrelevant in the RG sense, so it is not clear that a divergent Markov length must obstruct reversibility. 
Can we explicitly exhibit a class of states whose Markov length diverges but that are nevertheless connected to one another by a locally reversible path, and whose RG flow reaches a trivial state? How can these seemingly contradictory features be reconciled?

We have so far left the geometry of the CMI unspecified, and this hides a further subtlety. 
The Markov length sets the exponential decay of $I(A:C|B)$ with the width of the buffer $B$ in the \textit{local} tripartition, 
where at least one of $A$ and $C$ is microscopic~\cite{sang2025mixed} [see Fig.\ref{fig:cmi_geometries}(b)].
This is not the geometry in which the CMI or its scaling function was shown to decrease monotonically in Ref.~\cite{chen2026constraints}, where $A$ and $C$ are both much larger than $B$, the \textit{global} tripartition [see Fig.\ref{fig:cmi_geometries}(c)]. 
What, then, is the relation between the two? 
The distinction may shed light on the stability of mixed-state phases.
In particular, Ref.~\cite{lessa2024strong} showed that a state violating the local Markov property, such as one in a strong-to-weak symmetry breaking (SWSSB) phase~\cite{lee2023quantum,lessa2024strong}, is unstable under strong-symmetry-breaking perturbations.
It is then natural to ask whether the global property is tied to stability in the same way.
Is there a {stable} phase that breaks the global Markov property while remaining locally Markovian?
\footnote{
The reverse question does not arise, as CMI can only grow when $A$ and $C$ are enlarged at fixed $B$ due to SSA.
}

To address these questions cleanly, we work with hierarchical models, for which
the RG can be implemented exactly rather than approximately.  
This strategy is inspired by Dyson, who used a hierarchical Ising model to establish a phase
transition in a one-dimensional ferromagnet with long-range
interactions~\cite{dyson1969existence}.
The hierarchical construction also has a pure-state
counterpart in the exactly coarse-grainable tensor networks, such as tree tensor
networks and MERA~\cite{shi2006classical,vidal2008class}.  
Our main results are:

(a) For a mixed quantum state $\rho$, we show that the CMI between nonadjacent regions $A$ and $C$ conditioned on its buffer region $B$ is UV-finite, that is, bounded
in the scaling limit at fixed macroscopic geometry, whenever there \emph{exists} a locally reversible
RG which keeps the on-site Hilbert space dimension
fixed at every scale (Sec.\ref{sec:UV_finite}). 
We make this requirement precise in Sec.\ref{sec:RG_framework}, following Ref.~\cite{sang2024mixed}. 
We emphasize that the criterion does not require knowing how to implement the RG scheme, only that such a scheme exists.
This answers, in one direction, the question of how UV-finiteness is related to
the RG.  

(b) We present a majority-rule variant of Dyson's hierarchical Ising model whose Markov length diverges at every temperature $0<T<\infty$  (Sec.~\ref{sec:majority_rule_hierarchical_ising}).  
A locally-reversible RG can
nevertheless be implemented exactly at each scale, and it drives
the system toward the trivial infinite-temperature product state.
This realizes a class of states that have a divergent Markov length and yet are two-way connected to one another.
We show that these states can be connected in both directions to the
infinite-temperature product state if circuits with \emph{sublinear} range are
allowed \cite{sang2024mixed}.
However, under the refined definition of Ref.~\cite{sang2025mixed}, which
restricts circuits to \emph{polylogarithmic} range, they are not phase-equivalent to a
trivial product state.
 
(c) 
We then present another model, which, at the zero-noise fixed point, breaks the  
global Markov property while respecting the local one (Sec.~\ref{sec:stable_tree}).  
This model is the classical
counterpart of the Bell tree of
Refs.~\cite{yadavalli2025noisy,sommers2025dynamically} and belongs to a long
tradition of models that repeatedly encode each parent spin into its children
on a tree~\cite{forney1966concatenated,evans2000broadcasting}. 
Interestingly, the fixed point can be understood as a \textit{concatenation} of two familiar fixed points, classical SSB and SWSSB, arranged on a tree.   
Like SWSSB, this model has a strong $\mathbb{Z}_2$ symmetry.
However, it is a distinct fixed point from SWSSB. In particular, while SWSSB is destabilized by channels that {explicitly} break the strong symmetry, we show that the fixed point is stable against such perturbations.
In Sec.~\ref{sec:tree_cmi}, we further increase the error strength and give evidence that the CMI decays polynomially at the threshold, although the corresponding mutual information vanishes identically.

We conclude in Sec.~\ref{sec:conclusion} by discussing several directions for future work.  Technical details are collected in the appendices.

\tableofcontents

\section{Renormalization group and conditional mutual information} 
\label{sec:RG_framework}

We begin by viewing a real-space RG transformation as a sequence of
coarse-graining maps
\begin{equation}
\rho=\rho^{(0)} \xrightarrow{\mathcal R^{(0)}} \rho^{(1)}
\xrightarrow{\mathcal R^{(1)}} \cdots
\xrightarrow{\mathcal R^{(r-1)}} \rho^{(r)}
\xrightarrow{\mathcal R^{(r)}}\cdots ,
\end{equation}
where each $\mathcal R^{(r)}$ is a finite-depth circuit of local channels whose
range is uniformly bounded in units of the lattice at that RG scale.  Each
step reduces linear system sizes by a factor $b>1$. 

\begin{figure}[H]
\centering
\includegraphics[width=\linewidth]{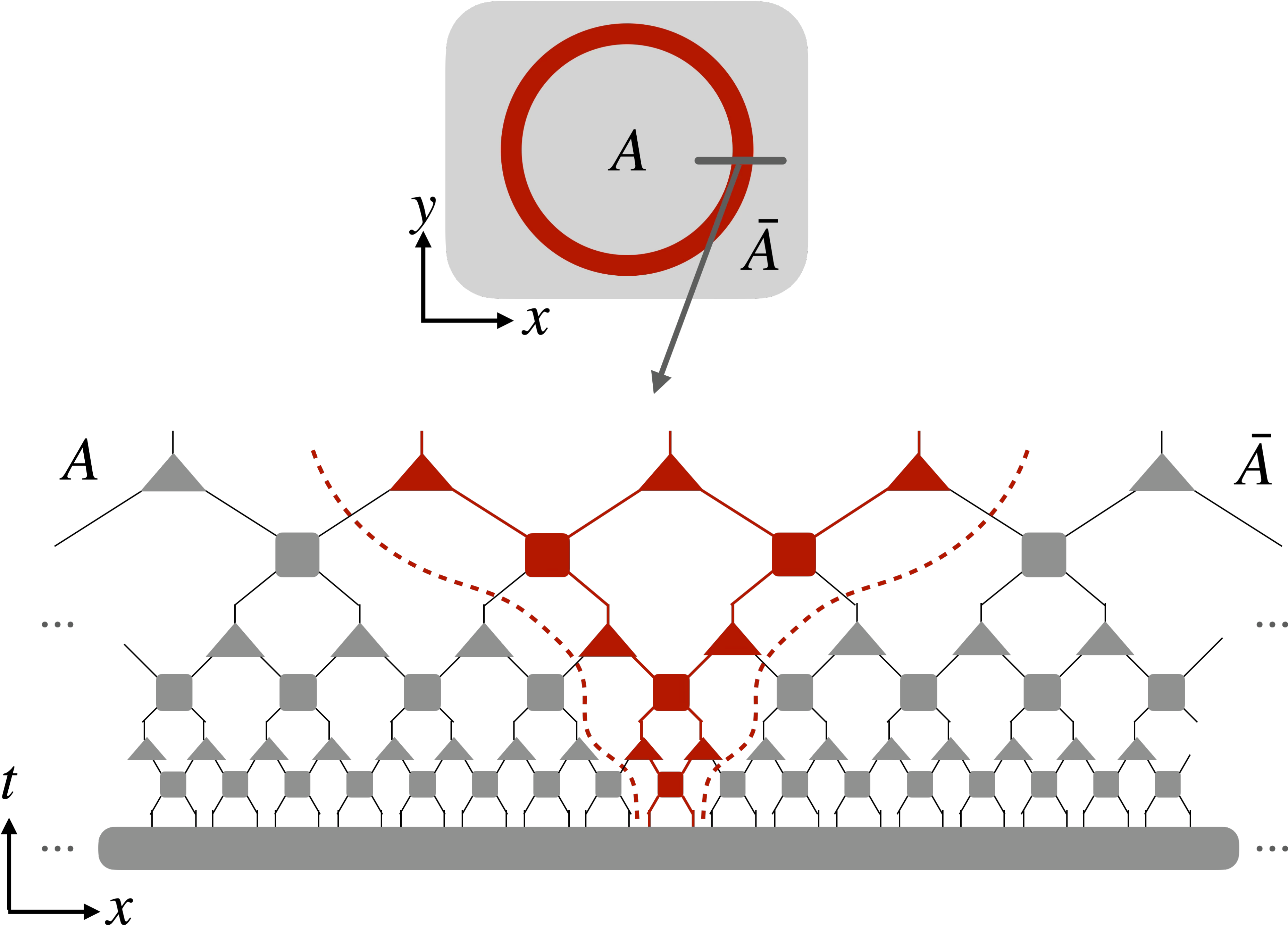}
\caption{Bounded causal tube in a real-space RG.  
The upper
schematic shows a fixed-width collar (red) around the boundary of $A$ in two
dimensions.  
The lower schematic shows a one-dimensional cut through the
$A|\bar A$ interface in a binary ($b=2$) MERA-like architecture.  
Since every horizontal cut intersects only
$O(1)$ red bonds, the tube has uniformly bounded transverse width.}
\label{fig:causal} 
\end{figure}

We emphasize that the RG circuit is local only at each scale, rather than
strictly local with respect to the original lattice.  This
scale-local structure implies the usual \textit{bounded causal-cone} property of
real-space RG, familiar from pure-state schemes such as
MERA~\cite{vidal2008class}.  As we show in Sec.~\ref{sec:UV_finite}, this
property is essential for proving our main result there.
Consider a collar of fixed microscopic width
around the boundary of a region $A$; the upper part of Fig.~\ref{fig:causal}
shows a two-dimensional example with $A$ a disk.  A single RG layer can enlarge
the collar by at most $O(1)$ sites, whereas blocking reduces its width by the
factor $b$.  After each layer, its width is therefore bounded by the previous
width divided by $b$, plus an $O(1)$ correction. 
Therefore, the causal
closure of the boundary collar forms a multiscale tube of uniformly bounded
transverse width.  The lower part of Fig.~\ref{fig:causal} illustrates such a
tube in a one-dimensional cut through a binary ($b=2$) MERA architecture.

So far, we have imposed no condition on the individual local channels
composing $\mathcal R^{(r)}$.  A natural requirement for a physically
reasonable RG scheme is that each such channel be
\textit{correlation-preserving}~\cite{sang2024mixed}.
Specifically, a channel $R_{X'\leftarrow X}$ maps a region $X$ to a
coarse-grained region $X'$ while preserving mutual information with its
complement:
\begin{equation}
\label{Eq:correlation_preserving}
I_{\rho'}(X':\bar{X})=I_{\rho}(X:\bar{X}),
\end{equation}
where $\rho'=R_{X'\leftarrow X}[\rho]$.
This criterion ensures that coarse graining loses no correlations between $X$
and $\bar X$, while allowing information internal to $X$ and uncorrelated with
$\bar X$ to be discarded.

There are many mixed states that can be coarse-grained using correlation-preserving channels.
A standard example is Kadanoff block-spin decimation for a
one-dimensional nearest-neighbor Ising chain~\cite{kadanoff1966scaling,
sang2024mixed}.  
Exact pure-state hierarchical constructions, such as
MERA~\cite{vidal2008class} and tree tensor networks~\cite{shi2006classical},
provide another class of examples.
A genuinely mixed quantum example is the
exact correlation-preserving RG for the finite-temperature two-dimensional
toric-code Gibbs state constructed in Ref.~\cite{sang2024mixed}. 
When exact
preservation is unavailable, the same guiding principle, namely retaining long-range correlations, is implicit in practical RG schemes~\cite{White92,vidal2007entanglement,levin2007tensor,gu2008tensor,evenbly2015tensor}.  See also Refs.~\cite{koch2018mutual,hu2020machine}, where RG transformations are learned precisely by minimizing the violation of correlation preservation.

{ 
A closely related notion is $s$-sourcery~\cite{swingle2016constructions,swingle2016},
which counts the copies of a state at linear size $L$ needed to build it at size
$2L$. 
Restricting to a single copy, Ref.~\cite{swingle2016} calls a state a
\emph{noisy} fixed point if $\rho_{2L}=V(\rho_L\otimes\sigma)V^{\dagger}$, with
$V$ a quasilocal unitary and $\sigma$ a product state, and an \emph{open} one if
$\rho_{2L}=\mathcal N(\rho_L)$ with $\mathcal N$ a quasilocal channel.
Correlation preservation is weaker than the noisy condition, which requires the discarded degrees of freedom to be uncorrelated
with everything rather than with $\bar X$ alone at fixed $X'$, and stronger than
the open condition, which requires no recovery map.  
} 

As discussed in Refs.~\cite{sang2024mixed}, the equality in
Eq.~\eqref{Eq:correlation_preserving} is equivalent to the existence of a local ``fine-graining'' channel
$\mathcal D_{X\leftarrow X'}$ satisfying
\begin{equation}
\left[
\left(\mathcal D_{X\leftarrow X'}\circ R_{X'\leftarrow X}\right)
\otimes \operatorname{id}_{\bar X}
\right](\rho)=\rho.
\label{eq:exact_recovery} 
\end{equation} 
In other words, each coarse-graining step can be undone by a channel  acting only
on the coarse-grained region: the step is locally reversible on $\rho$.   
We emphasize that $\mathcal{D}_{X \leftarrow X'}$ acts as an inverse of $\mathcal{R}_{X'\leftarrow X}$ only for the particular state $\rho$, as a channel generally does not have an inverse. 
Following the terminology
of Ref.~\cite{sang2024mixed}, we call an RG scheme built entirely from such
gates an \textit{ideal} RG scheme.

We note that, for exact isometric pure-state RG constructions such as MERA and TTN, the
fine-graining channel $\mathcal D_{X\leftarrow X'}$ exists by construction.  This property is not guaranteed for general
mixed-state RG schemes, even when they have the same architecture as MERA or TTN.  

\subsection{Ideal RG implies conditional mutual information is UV-finite}
\label{sec:UV_finite}

\begin{figure*}[t] 
\centering 
\includegraphics[width=\linewidth]{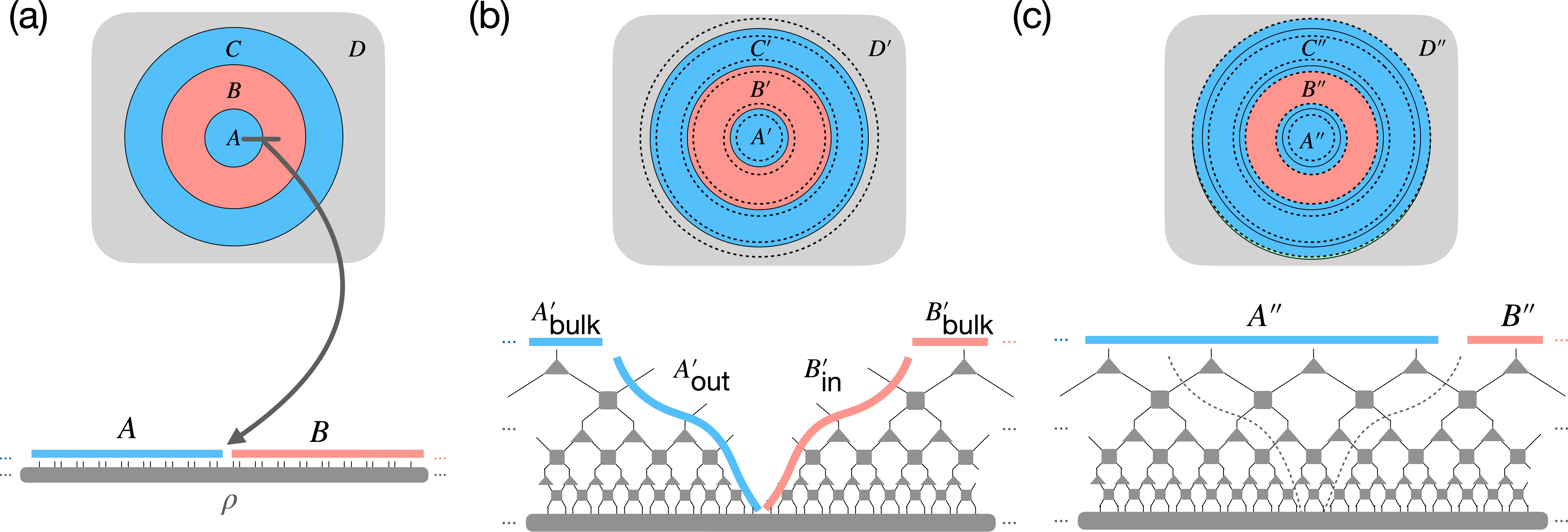}
\caption{The three steps of the proof of UV-finite CMI.
(a) The annular tripartition $A|B|C$, with $D$ the complement.  The lower
schematic shows a one-dimensional cut through the $A|B$ interface.
(b) Step 1: applying the RG gates outside the interface coarse-grains the
bulk and leaves the interface degrees of freedom untouched, without
changing the CMI.  
The dashed curves show the boundaries of the tubes.  
(c) Steps 2
and 3: once the two sides of each interface are grouped together, the remaining
interface gates can be applied, and the relevant
systems reduce to finite coarse regions $A'',B'',C''$, with $D''$ the remaining
complement.  The lower schematics in (b) and (c) show a representative $A|B$
interface, and the $B|C$ and $C|D$ interfaces are treated analogously.   
}
\label{fig:rg_cmi_finite}   
\end{figure*}  

We now present one of our main results: for a mixed state $\rho$ in $d$ spatial dimensions, the CMI is UV-finite whenever there exists an ideal RG with a fixed finite Hilbert space dimension $d_*$ at every scale.  
By UV-finite, we mean that $I(A:C|B)$ remains uniformly bounded in the lattice 
scaling limit
\begin{equation}
\frac{N_A}{N_B}\to x_A,
\quad
\frac{N_C}{N_B}\to x_C,
\quad
N_B\to\infty, 
\label{eq:UV_scaling}
\end{equation} 
where $0<x_A,x_C<\infty$, $N_X$ is the number of lattice sites in region
$X$, and $d$ is the spatial dimension. 
We also require that $A$ and $C$ never come close to each other, by assuming
that $\operatorname{dist}(A,C)$ is bounded below by a fixed positive fraction of
the linear size $N_B^{1/d}$ of $B$. 

For concreteness, we prove the bound for a sequence of increasingly large
annular tripartitions with the same fixed shape, as shown in
Fig.~\ref{fig:rg_cmi_finite}(a).  Here $A$ is the inner region, $B$ is an
annular buffer, $C$ lies outside the buffer, and $D$ is the complement.  The
same argument applies to other fixed-shape sequences for which $B$ separates $A$ and $C$.
  
\paragraph*{Proof.}
We will proceed in three steps, which we briefly summarize the idea as follows (see
Fig.~\ref{fig:rg_cmi_finite} for illustration).
We first stop the RG at a fixed scale and coarse-grain everything \emph{outside} the interface causal tubes, which leaves the CMI
unchanged.  
We then enlarge $A$ and $C$ so that each tube lies inside a single
region, and strong subadditivity (SSA) implies that this enlargement can only increase the CMI.
Finally, we complete the RG inside the tubes and bound what remains by its dimension.
 
Let $b$ be the RG block size, and index the fixed-shape sequence by $m$ so
that $N_X(m)=\Theta(b^{dm})$ for $X=A,B,C$ (recall that $d$ denotes the spatial dimension).  
In other words, $m$ is roughly the logarithmic size parameter, and the regions admit $m$ RG steps before
reaching an $O(1)$ size.  
We choose a fixed $q$,
independent of $m$, such that after $(m-q)$ RG steps the coarse-grained geometry has
$O(1)$ size while the causal tubes around adjacent interfaces remain disjoint. 
Our goal is to bound the CMI by a constant independent of $m$ (although it may depend on $b$, $d$, and $d_*$).

\emph{Step 1: CMI-preserving bulk coarse-graining.---} From the full RG circuit formed
by these $(m-q)$ layers, we construct a \emph{partial circuit} by removing every gate
whose support meets one of the tubes around the $A|B$, $B|C$, and $C|D$
interfaces.  
This gives a partially coarse-grained
state $\rho'$ shown in
Fig.~\ref{fig:rg_cmi_finite}(b), with region factors $A',B',C',D'$.
By construction, the gates belonging to the partial circuit lie wholly inside one of $A,B,C,D$. 
  
We now show CMI preservation directly.   
The essential point is that every gate belonging to the partial circuit remains correlation preserving on the state it applies to.
We note that since correlation preservation is a property of both the channel and the state, this needs to be verified more carefully, as we show in App.~\ref{app:modified_circuit_recovery}.  
Furthermore, the corresponding recovery channel acts only on the support of the gate and therefore commutes with tracing out the rest, so exact recovery also holds on every marginal containing that support. 
Therefore, using
$I(A:C|B)=I(A:BC)-I(A:B)$, one finds that every gate in the partial circuit preserves the CMI, and thus 
\begin{equation}
I_{\rho}(A:C|B)=I_{\rho'}(A':C'|B').
\label{eq:bulk_cleaning_cmi}
\end{equation} 

\emph{Step 2: Causal tube enlargement by SSA.---}
Since the previously removed gates act on both sides, they need not preserve
$I_{\rho'}(A':C'|B')$.
To circumvent this, we regroup the systems so that each tube lies inside a single region.  To do so, 
we decompose the primed systems as
\begin{equation}
\begin{aligned}
A'&=A'_{\mathrm{bulk}}A'_{\mathrm{out}},
&\qquad
B'&=B'_{\mathrm{in}}B'_{\mathrm{bulk}}B'_{\mathrm{out}},
\\
C'&=C'_{\mathrm{in}}C'_{\mathrm{bulk}}C'_{\mathrm{out}},
&\qquad
D'&=D'_{\mathrm{in}}D'_{\mathrm{bulk}}.
\end{aligned}
\label{eq:hybrid_region_decomposition}
\end{equation}
Here, the bulk factors live at the coarse-grained scale, while the in/out factors contain
the interface degrees of freedom left over at all scales.   
The lower part of Fig.~\ref{fig:rg_cmi_finite}(b) illustrates the split at the
$A|B$ interface. 

A straightforward application of the chain rule and strong subadditivity gives
\begin{equation}
I(A':C'|B')
\leq
I\!\left(
A'B'_{\mathrm{in}}:
B'_{\mathrm{out}}C'D'_{\mathrm{in}}
\middle|
B'_{\mathrm{bulk}}
\right).
\label{eq:separator_enlargement}
\end{equation}
Indeed, the difference between the right- and left-hand sides, i.e., 
$I\!\left(
A'B'_{\mathrm{in}}:
B'_{\mathrm{out}}C'D'_{\mathrm{in}}
\middle|
B'_{\mathrm{bulk}}
\right)-I(A':C'|B')$, is the sum of the three CMIs
$I\!\left(
B'_{\mathrm{in}}:B'_{\mathrm{out}}C'D'_{\mathrm{in}}
\middle|B'_{\mathrm{bulk}}
\right)$, $I\!\left(
A':B'_{\mathrm{out}}
\middle|B'_{\mathrm{in}}B'_{\mathrm{bulk}}
\right)$, and $I\!\left(
A':D'_{\mathrm{in}}
\middle|B'C'
\right)$, and is therefore nonnegative by SSA.

\emph{Step 3: Completion of the interface tubes.---} We now bound the
right-hand side of Eq.~\eqref{eq:separator_enlargement}.  
We group
$A'_{\mathrm{out}}B'_{\mathrm{in}}$ on the left,
$B'_{\mathrm{out}}C'_{\mathrm{in}}$ and
$C'_{\mathrm{out}}D'_{\mathrm{in}}$ on the right, and then apply the RG gates that lie on the interface.
This gives the channels illustrated in
Fig.~\ref{fig:rg_cmi_finite}(c):
\begin{equation}
\begin{aligned}
A'B'_{\mathrm{in}}&\rightarrow A'',
\quad
B''\equiv B'_{\mathrm{bulk}},\quad
B'_{\mathrm{out}}C'D'_{\mathrm{in}}\rightarrow C''.
\end{aligned}
\label{eq:tube_side_channels}
\end{equation}
The factor $D'_{\mathrm{bulk}}$ belongs to the remaining complement and is
traced out.
We denote the output state of the side channels by $\rho''$.
Furthermore, each gate also remains correlation preserving on the state it now acts on due to exactly the same reasoning that leads to Eq.~\eqref{eq:bulk_cleaning_cmi}.
Repeating the
argument of Step 1 therefore gives
\begin{equation}
I_{\rho'}\!\left(
A'B'_{\mathrm{in}}:
B'_{\mathrm{out}}C'D'_{\mathrm{in}}
\middle|
B'_{\mathrm{bulk}}
\right)
=
I_{\rho''}(A'':C''|B'').
\label{eq:tube_cmi_equality}
\end{equation}

Crucially, due to the causal-cone property of
Sec.~\ref{sec:RG_framework} (Fig.~\ref{fig:causal}), $A''$ and $C''$ each contain at most $n_*=O(1)$ sites, where $n_*$
is independent of $m$.  Therefore,
\begin{equation}
\begin{aligned}
I_{\rho''}(A'':C''|B'')
&\leq
2\min\{\log\dim A'',\log\dim C''\}
\\
&\leq
2n_*\log d_*.
\end{aligned}
\label{eq:finite_separator_bound}
\end{equation}
Combining Eqs.~\eqref{eq:bulk_cleaning_cmi},
\eqref{eq:separator_enlargement}, \eqref{eq:tube_cmi_equality}, and
\eqref{eq:finite_separator_bound} proves that $I_{\rho}(A:C|B)$ is bounded by
a constant independent of the logarithmic size parameter $m$.  

\begin{figure*}[!t]
\centering
\begin{minipage}[t]{0.31\textwidth}
\centering
\begin{tikzpicture}[x=0.85cm,y=0.62cm,font=\footnotesize]
\def\h{0.48}
\draw[fill=regionAC] (0,0) rectangle (1.8,\h);
\draw[fill=regionB]  (1.8,0) rectangle (3.6,\h);
\draw[fill=regionAC] (3.6,0) rectangle (5.4,\h);
\draw[dashed,gray] (2.7,-0.25) -- (2.7,\h+0.35);
\node at (0.9,0.5*\h) {$A$};
\node at (2.7,0.5*\h) {$B$};
\node at (4.5,0.5*\h) {$C$};
\node[gray,below] at (2.7,-0.25) {center};
\end{tikzpicture}
\smallskip

(a) All-region-extensive: $r_N\to r_0$
\end{minipage}
\hfill
\begin{minipage}[t]{0.31\textwidth}
\centering
\begin{tikzpicture}[x=0.85cm,y=0.62cm,font=\footnotesize]
\def\h{0.48}
\draw[fill=regionAC] (0,0) rectangle (0.35,\h);
\draw[fill=regionB]  (0.35,0) rectangle (5.05,\h);
\draw[fill=regionAC] (5.05,0) rectangle (5.4,\h);
\draw[dashed,gray] (2.7,-0.25) -- (2.7,\h+0.35);
\node[above] at (0.175,\h) {$A$};
\node at (2.7,0.5*\h) {$B$};
\node[above] at (5.225,\h) {$C$};
\node[gray,below] at (2.7,-0.25) {center};
\end{tikzpicture}
\smallskip

(b) Local: $r_N\to0$
\end{minipage}
\hfill
\begin{minipage}[t]{0.31\textwidth}
\centering
\begin{tikzpicture}[x=0.85cm,y=0.62cm,font=\footnotesize]
\def\h{0.48}
\draw[fill=regionAC] (0,0) rectangle (2.55,\h);
\draw[fill=regionB]  (2.55,0) rectangle (2.85,\h);
\draw[fill=regionAC] (2.85,0) rectangle (5.4,\h);
\draw[dashed,gray] (2.7,-0.25) -- (2.7,\h+0.35);
\node at (1.275,0.5*\h) {$A$};
\node[above] at (2.7,\h) {$B$};
\node at (4.125,0.5*\h) {$C$};
\node[gray,below] at (2.7,-0.25) {center};
\end{tikzpicture}
\smallskip

(c) Global (thin-buffer): $r_N\to\infty$
\end{minipage}
\caption{Three tripartitions distinguished by the asymptotic ratio
$r_N=N_A/N_B$, with $N_A=N_C$.  The microscopic regions in (b) and (c) are
drawn wider than their asymptotic relative sizes for  visibility.}
\label{fig:cmi_geometries}
\end{figure*}
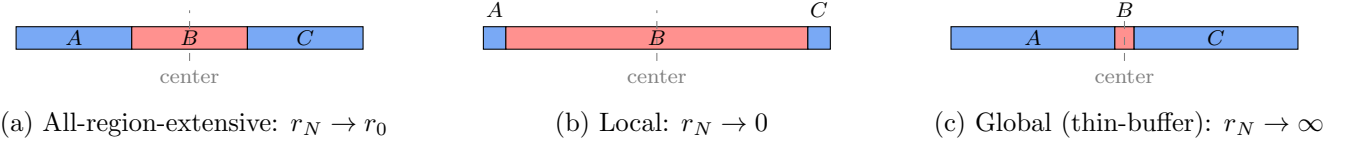 
 
\subsection{Combining with the scaling relation}
\label{sec:rg_scaling} 
One of the most powerful aspects of RG is that it relates the description of a
fixed microscopic system at different length scales to a flow through parameter
space.  Specifically, if
$\{t_i\}$ are scaling fields with RG exponents $\{1/\nu_i\}$, then
\begin{equation}
\label{Eq:scaling_hypothesis}
\rho^{(r+1)}_X(\{t_i\})=\rho^{(r)}_X(\{b^{1/\nu_i}t_i\}),
\end{equation}
where $\rho_X^{(r)}$ is the reduced state of a fixed rescaled region $X$ at RG
scale $r$.\footnote{An RG transformation consists of coarse-graining followed
by rescaling, so the two sides can be compared on the same Hilbert space $X$.}

The previous subsection proves that an ideal RG with a finite local Hilbert-space dimension implies that the CMI is UV-finite.
While UV finiteness alone does not necessarily guarantee convergence to a unique
scaling function, it is natural to expect that boundedness allows one to choose a
suitable fixed-shape family along which the scaling limit exists.
The CMI then has the asymptotic form
\begin{equation}
\mathrm{CMI}_{\rho^{}(\{t_i\})}(\vec N)
=f\!\left(\frac{N_A}{N_B},\frac{N_C}{N_B},
\{t_iN_B^{1/(\nu_i d)}\}\right)+o(1).
\label {eq:cmi_scaling_function}
\end{equation} 
Strictly speaking, the discrete RG transformation allows the scaling function
to retain a period-one dependence on $\log_b(N_B^{1/d})$.  We therefore
restrict to $N_B=b^{ds}$, with $s$ a nonnegative integer, so that the phase of
this periodic dependence is fixed.  As an aside, we note that this dependence
disappears in the $b\rightarrow 1^+$ limit.
  
We close with a simple interpretation of Eq.~\eqref{eq:cmi_scaling_function} that will be useful for the example in
Sec.~\ref{sec:majority_rule_hierarchical_ising}: in the scaling limit, the CMI is \textit{invariant} under an RG step,
provided all three regions are rescaled together.
In fact, we now show that if the CMI is exactly invariant under each RG
step, then Eq.~\eqref{eq:cmi_scaling_function} holds exactly, without the
finite-size $o(1)$ correction.  
To begin with, exact RG invariance implies
\begin{equation}
\label{Eq:cmi_invariant_trees}
\mathrm{CMI}_{\rho^{(r)}(\{t_i\})}(\vec N)
=
\mathrm{CMI}_{\rho^{(r+1)}(\{t_i\})}(\vec N/b^d).
\end{equation}
Combining this identity with Eq.~\eqref{Eq:scaling_hypothesis} and iterating
$s$ times for
$N_B=b^{ds}$ gives
\begin{equation}
\label{Eq:CMI_uvfinite_main}
\mathrm{CMI}_{\rho^{(0)}(\{t_i\})}(\vec N)
=
\mathrm{CMI}_{\rho^{(0)}(\{t_iN_B^{1/(\nu_i d)}\})}
\!\left(\frac{N_A}{N_B},1,\frac{N_C}{N_B}\right),
\end{equation}
which is the desired result.

\subsection{Three asymptotic CMI geometries}
\label{sec:cmi_geometries}

Before ending this section, we discuss the three asymptotic geometry classes illustrated in
Fig.~\ref{fig:cmi_geometries}, which will be used in the examples below. 
We let the interval $ABC$ span the whole system, with the middle region $B$ having length $N_B$ and the two side
regions having equal length $N_A=N_C=(N-N_B)/2$. 
The CMI geometry is then distinguished by $r_N=N_A/N_B$ in the scaling limit. 
In particular, we consider
\begin{itemize}
\item \textit{All-region-extensive geometry}: $r_N\to r_0\in(0,\infty)$,
so $A$, $B$, and $C$ occupy fixed nonzero fractions of the system.  
\item \textit{Local geometry}: $r_N\to0$; we take both $A$ and $C$
microscopic while $B$ grows.  The decay of the CMI with $N_B$ defines the
Markov length. 
\footnote{
We note that our definition is a special case of the one in
Ref.~\cite{sang2025stability}, which requires
$I(A:C|B)\leq\mathrm{poly}(|A|,|C|)\,e^{-\operatorname{dist}(A,C)/\xi}$ for
\emph{every} simply connected region $A$, every annular buffer $B$ around it,
and $C$ the complement of $A\cup B$.  We impose this decay only along the
single family of shapes above, in which $|A|$ and $|C|$ are $O(1)$ and the
prefactor is a constant.  A divergence here therefore forces a divergence in
their sense.  
}
\item \textit{Global (thin-buffer) geometry}: $r_N\to\infty$; $A$ and $C$
are macroscopic, whereas $B$ is subextensive and often microscopic.  This
limit probes global information that the thin buffer fails to 
screen.  
Furthermore, SSA is more constraining here.  Writing the UV-finite global CMI as $F(x)$ with
$x=tN_B^{1/(\nu d)}$, enlarging the buffer by one RG step
gives~\cite{chen2026constraints}
\begin{equation}
F\!\left(b^{1/\nu}x\right)\le F(x).
\label{eq:global_cmi_scaling_monotonicity}
\end{equation} 
Thus $F$ is nonincreasing along the sequence
$x,\,b^{1/\nu}x,\,b^{2/\nu}x,\ldots$, whose spacing is set by the blocking
factor $b$.  In the $b\to1^{+}$ limit the CMI 
is then a monotonically decreasing function of
$x$~\cite{chen2026constraints}.
\end{itemize}

\section{Majority-rule hierarchical Ising model} 
\label{sec:majority_rule_hierarchical_ising}

We now present a simple classical mixed-state example with a divergent Markov length at every nonzero finite temperature.  Nevertheless, an ideal RG can be implemented exactly and suggests that the system flows to the trivial
infinite-temperature product state.  
The model is a natural majority-rule variant of
Dyson's hierarchical Ising model~\cite{dyson1969existence}.  
To our knowledge, this variant has not been studied before.

Consider $N=3^n$ Ising spins 
$s_j=\pm 1$, $j=1,\ldots,N$.  The spin variables at level $p$ are defined
recursively by 
\begin{equation}
\label{eq:maj_hierarchical_spins}
s_j^{(0)}=s_j,\qquad
s_j^{(p+1)}=\operatorname{maj}\!\left(
s_{3j-2}^{(p)},s_{3j-1}^{(p)},s_{3j}^{(p)}\right),
\end{equation}
where $\operatorname{maj}(a,b,c)={(a+b+c-abc)}/{2}.
$
The Hamiltonian is
\begin{equation}
\label{eq:maj_hierarchical_hamiltonian}
\begin{aligned}
H_N & =
-\sum_{p=0}^{n-1}3^{-p/\nu}
H^{(p)}(\{s_j\}),\\
H^{(p)} & = \sum_{j=1}^{N/3^{p+1}}
\left(
s_{3j-2}^{(p)}s_{3j-1}^{(p)}
{}+s_{3j-1}^{(p)}s_{3j}^{(p)}
{}+s_{3j}^{(p)}s_{3j-2}^{(p)}
\right).
\end{aligned}
\end{equation}
Intuitively, $H^{(p)}$ is the three-site Ising Hamiltonian for the
level-$p$ hierarchical spins.
 
\begin{figure}[H]
\centering
\begin{tikzpicture}[
  x=0.645cm,
  y=0.78cm,
  font=\footnotesize,
  microscopic/.style={
    circle, draw=blue!55!black, fill=blue!8,
    minimum size=5.2mm, inner sep=0.5pt
  },
  coarse/.style={
    rounded corners=1.5pt, draw=red!60!black, fill=red!8,
    minimum width=8mm, minimum height=5.2mm, inner sep=1pt
  },
  root spin/.style={
    rounded corners=1.5pt, draw=green!45!black, fill=green!10,
    minimum width=8mm, minimum height=5.2mm, inner sep=1pt
  },
  majority gate/.style={
    rounded corners=1pt, draw=gray!65, fill=white,
    inner xsep=2pt, inner ysep=1pt, font=\scriptsize
  },
  rg feed/.style={
    draw=gray!75, line width=0.55pt,
    preaction={draw=white, line width=2.2pt}
  },
  rg arrow/.style={
    rg feed, -{Latex[length=1.6mm,width=1.05mm]},
    shorten <=2.5mm, shorten >=3mm
  },
  coupling/.style={draw=orange!80!black, line width=0.9pt}
]

\foreach \i in {1,...,9}
  \coordinate (s\i) at (\i-1,0);
\coordinate (u1) at (1,2.3);
\coordinate (u2) at (4,2.3);
\coordinate (u3) at (7,2.3);
\coordinate (r)  at (4,4.4);
\coordinate (g1) at (1,1.1);
\coordinate (g2) at (4,1.1);
\coordinate (g3) at (7,1.1);
\coordinate (g4) at (4,3.35);
 
\foreach \a/\b/\c in {1/2/3,4/5/6,7/8/9}{
  \draw[coupling] (s\a) to[bend left=35] (s\b);
  \draw[coupling] (s\b) to[bend left=35] (s\c);
  \draw[coupling] (s\a)
    .. controls +(0,-0.8) and +(0,-0.8) .. (s\c);
}
\draw[coupling] (u1) to[bend left=12] (u2);
\draw[coupling] (u2) to[bend left=12] (u3);
\draw[coupling] (u1)
  .. controls +(0,-0.8) and +(0,-0.8) .. (u3);

\foreach \a/\b/\c/\g/\u in
  {1/2/3/g1/u1,4/5/6/g2/u2,7/8/9/g3/u3}{
  \draw[rg feed] (s\a) -- (\g);
  \draw[rg feed] (s\b) -- (\g);
  \draw[rg feed] (s\c) -- (\g);
  \draw[rg arrow] (\g) -- (\u);
}
\foreach \u in {u1,u2,u3}
  \draw[rg feed] (\u) -- (g4);
\draw[rg arrow] (g4) -- (r);

\foreach \i in {1,...,9}
  \node[microscopic] at (s\i) {$s_{\i}$};
\node[coarse] at (u1) {$s_1^{(1)}$};
\node[coarse] at (u2) {$s_2^{(1)}$};
\node[coarse] at (u3) {$s_3^{(1)}$};
\node[root spin] at (r) {$s_1^{(2)}$};
\foreach \g in {g1,g2,g3,g4}
  \node[majority gate] at (\g) {$\operatorname{maj}$};

\node[anchor=east, gray!70!black] at (-0.65,0) {$p=0$};
\node[anchor=east, gray!70!black] at (-0.65,2.3) {$p=1$};
\node[anchor=east, gray!70!black] at (-0.65,4.4) {$p=2$};
\node[anchor=west, orange!55!black] at (8.55,0) {$J_0=1$};
\node[anchor=west, orange!55!black] at (8.55,2.3)
  {$J_1=3^{-1/\nu}$};
 
\draw[gray!75, -{Latex[length=1.6mm,width=1.05mm]}]
  (0.1,-1.05) -- (0.9,-1.05);
\node[anchor=west] at (1.05,-1.05) {majority map};
\draw[coupling] (4.7,-1.05) -- (5.5,-1.05);
\node[anchor=west] at (5.65,-1.05) {coupling in $H^{(p)}$}; 
\end{tikzpicture}
\caption{First two steps of the majority hierarchy.  
Gray branches
feed each three-spin block through a majority gate to its coarse(majority) spin.  
Orange
arcs show the three pairwise couplings within every level-$p$ block, with
strength $J_p=3^{-p/\nu}$.
}
\label{fig:maj_hierarchical_ising}
\end{figure}
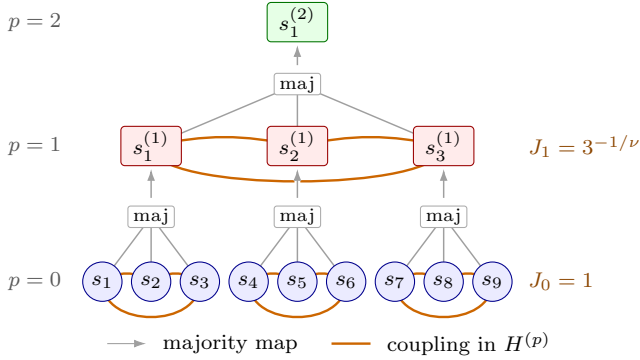

The corresponding Gibbs distribution is
\begin{equation}
\label{Eq:rho_maj_hierarchy}
\rho(T,N)(\{s_j\})=\frac{1}{Z(T,N)}
\exp\left[-\frac{1}{T}H_N(\{s_j\})\right].
\end{equation}
The first two levels of the construction are shown in
Fig.~\ref{fig:maj_hierarchical_ising}.
   
\subsection{Tension between the divergent 
Markov length and the ideal RG flow to a trivial state}
Let us now implement ideal RG for the mixed state in Eq.~\eqref{Eq:rho_maj_hierarchy}.
The most natural coarse-graining channel is to map every block of three spins to its majority spin,
\begin{equation}
\label{eq:maj_channel}
\left(\mathcal{E}_j\right)^{s'}_{a,b,c}
=\delta_{s',\operatorname{maj}(a,b,c)},\qquad
\mathcal{E}=\prod_j\mathcal{E}_j .
\end{equation} 
This channel is correlation preserving for the hierarchical Gibbs state, since the majority spin is precisely the variable through which a block couples to all longer-distance levels.  
Therefore, replacing a block by its
majority spin loses only intra-block information.   

We can now derive the {exact} relation between $\rho(T,N)$ and the coarse-grained state $\mathcal{E}[\rho(T,N)]$. 
The key point is that summing over the bottom spins $a,b,c = \pm 1$, subject to the majority constraint, does not influence the Hamiltonian in the higher hierarchy.  Therefore, computing $\mathcal{E}[\rho(T,N)]$ requires only the following identity
\begin{equation} 
\label{eq:app_block_weight}
\begin{aligned}
\sum_{a,b,c}
\delta_{s,\operatorname{maj}(a,b,c)}
\exp\left[(ab+bc+ca)/T\right] 
& = e^{3/T}+3e^{-1/T}, 
\end{aligned}
\end{equation}
which is independent of $s$. Therefore, applying the majority-vote channel to the lowest remaining
hierarchical layer gives only an overall normalization factor.
\footnote{This independence is what makes the
majority-rule model simpler than Dyson's original hierarchical Ising model~\cite{dyson1969existence}.}
 
Now, rewriting all higher
layers in terms of the majority spins, one finds that their couplings are shifted by one level: $\sum_{p=1}^{n-1}3^{-p/\nu}H^{(p)}
=3^{-1/\nu}
\sum_{p'=0}^{n-2}3^{-p'/\nu}H_{\mathrm{coarse}}^{(p')}.$
Here $H_{\mathrm{coarse}}^{(p')}$ is the same level-$p'$ interaction evaluated
on the majority spins after one coarse-graining step.
It follows that
\begin{equation}
\mathcal{E}\!\left[\rho(T,N)\right]
=\rho(T\,3^{1/\nu},N/3),
\end{equation}
and thus $T$ is a relevant scaling variable with RG eigenvalue $3^{1/\nu}$.  
The zero-temperature fixed point is then unstable, and finite temperature defines the crossover length $\xi=T^{-\nu}$.

We can also write the reverse channel explicitly.  For each coarse spin $s$,
define
\begin{equation}
\label{eq:maj_recovery_channel}
\left(\mathcal R_{T,j}\right)^{a,b,c}_{s}
=
\frac{\delta_{s,\operatorname{maj}(a,b,c)}
\exp\!\left[(ab+bc+ca)/T\right]}
{e^{3/T}+3e^{-1/T}}.
\end{equation}
Setting $\mathcal R_T=\prod_j\mathcal R_{T,j}$, one thus has $\mathcal R_T\!\left[\rho(T\,3^{1/\nu},N/3)\right]=\rho(T,N)$.
Therefore, the RG step is locally reversible
block by block in the sense of
Refs.~\cite{sang2024mixed,sang2025mixed}.\footnote{
The model is also an exact
noisy $s=1$ source fixed point in the sense of Ref.~\cite{swingle2016}. 
}
However, as we show in the following subsection, at fixed $\infty>T>0$ the CMI decays only algebraically as the buffer size $N_B\to\infty$, and hence the Markov length diverges.

{
Before calculating the CMI, let us first resolve the tension between the divergent 
Markov length and the RG flow to a trivial state.
Following Ref.~\cite{sang2025mixed}, a channel circuit is called local only if
both its gate range and its depth grow at most \textit{polylogarithmically} with the
system size.  
Naively, the polylogarithmic restriction looks like a minor technicality.  
One
might expect the weaker condition assumed in Ref.~\cite{sang2024mixed}, i.e., a
sublinear range $r = O(N^{a<1})$ with $r/N \to 0$, to be enough to
rule out a two-way connection between states with divergent and finite Markov
lengths.  
However, we now illustrate the importance of the polylogarithmic gate-range restriction. 
In particular, we will show that a circuit with range $r = O(N^{a<1})$
connects $\rho(T,N)$ at any fixed $0<T<\infty$ to the trivial infinite-temperature mixed state,
whereas a circuit with polylogarithmic range does not.
 
Specifically, write $\rho_k=\rho(T_k,N_k)$, where
$N_k=N/3^k$ and $T_k=T3^{k/\nu}$.  
We find that the relative entropy in the large-$T_k$, large-$N_k$ limit behaves as (see App.~\ref{app_sec:rel})
\begin{equation}
\label{eq:maj_rel_entropy}
\begin{split}
D \Big(\rho_k ||\frac{\mathbbm{1}}{2^{N_k}}\Big)
&\approx \frac{A_\nu}{2T^{2}}\,\frac{N}{3^{k(1+2/\nu)}},
\end{split}
\end{equation}
where $A_\nu={1}/{(1-3^{-(1+2/\nu)})}$. 
Therefore, $D(\rho_k\|\mathbbm{1}/2^{N_k})$ depends only on how $3^{k(1+2/\nu)}$ compares with $N$.

First take $k=\lceil a\log_3N\rceil$, the least integer at least $a\log_3N$, so
that $3^k=\Theta(N^a)$ with $a<1$.  Eq.~\eqref{eq:maj_rel_entropy} then gives
$D(\rho_k\|\mathbbm{1}/2^{N_k})=\Theta(N^{1-a(1+2/\nu)})$, which vanishes when
$a>\nu/(\nu+2)$.   
Therefore, for any
$\nu/(\nu+2)<a<1$, circuits of sublinear range connect the two states in both
directions at any fixed $T>0$.

Now restrict the gate range to $3^k=\Theta((\log N)^c)$ and thus $k=\Theta(\log\log
N)$, for which Eq.~\eqref{eq:maj_rel_entropy} gives
$D(\rho_k\|\mathbbm{1}/2^{N_k})=\Theta(N/(\log N)^{c(1+2/\nu)})\to\infty$.
Writing $\epsilon=\frac12\|\rho_k-\mathbbm{1}/2^{N_k}\|_1$, the sharp
entropy-continuity bound~\cite{fannes1973,audenaert2007sharp} gives
$D(\rho_k\|\mathbbm{1}/2^{N_k})\leq\epsilon N_k\log2+h_2(\epsilon)$, whose
right-hand side would vanish if the trace-norm difference decayed faster than
every inverse power of $N$.  This contradicts the divergence of the relative entropy.
Therefore, the ideal RG with  polylogarithmic range cannot map $\rho(0<T<\infty, N)$, to the infinite-temperature product state.

}

\subsection{Conditional mutual information}

We now compute the CMI for the majority-rule hierarchical Ising model.  We 
restrict to tripartitions in which the interval $ABC$ spans the whole
system of size $N=3^n$, the middle region $B$ has length $N_B$, and the two side
regions have equal length $N_A=N_C=(N-N_B)/2$.  We write the corresponding
conditional mutual information as $\mathrm{CMI}(T;N,N_B)=I(A:C|B)$.
 
As discussed in Sec.~\ref{sec:rg_scaling}, exact RG invariance gives an exact
discrete scaling identity without finite-size corrections whenever the 
boundaries between $A$, $B$, and $C$ coincide with RG block boundaries.
Specifically, for any block size $b=3^r$ that divides $N_A$, $N_B$, and
$N_C$,
\begin{equation}
\label{Eq:maj_UV_independent}
\mathrm{CMI}(T;N,N_B)
=\mathrm{CMI}(T b^{1/\nu};N/b,N_B/b).
\end{equation}
We apply Eq.~\eqref{Eq:maj_UV_independent} to three realizations of the
geometries in Fig.\ref{fig:cmi_geometries}: (i) All-region-extensive: $N_A=N_B=N/3$.
(ii) Local: $N_B$ is extensive and $N_A$ is subextensive.
(iii) Global: $N_B$ is subextensive and $N_A$ is extensive.

\subsubsection{All-region-extensive geometry.} 
\begin{figure}[H]
\centering
\includegraphics[width=\linewidth]{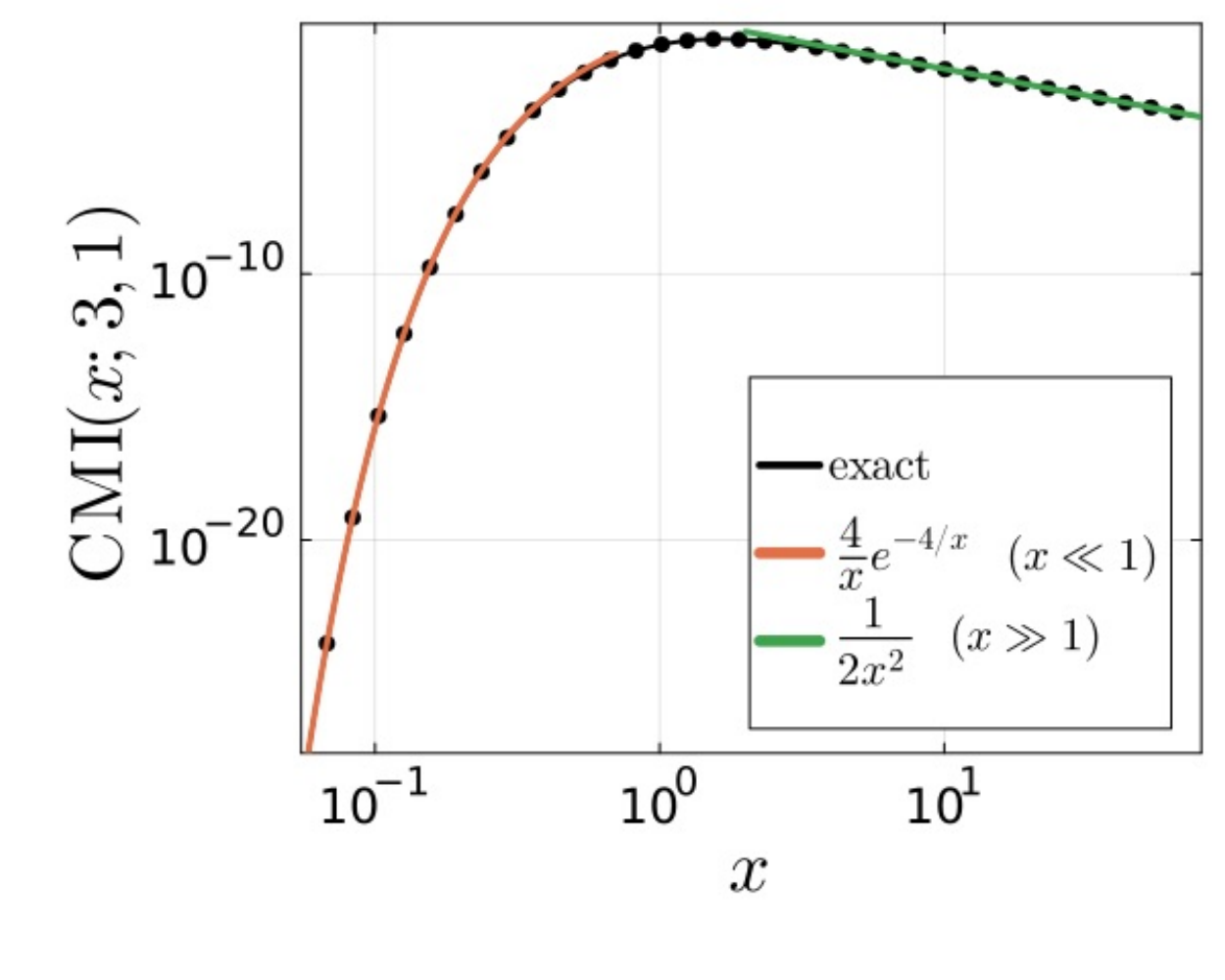}
\caption{
  The CMI of the majority-rule hierarchical model in the all-region-extensive geometry, i.e., $\mathrm{CMI}(T ;N,N/3) = \mathrm{CMI}(x;3,1)$, where $x =T N_B^{1/\nu}$.
}
\label{fig:maj_three_spin_cmi_limits}
\end{figure}

Choosing $b = N_B = N/3$, Eq.~\eqref{Eq:maj_UV_independent} implies 
$\mathrm{CMI}(T ;N,N/3) = \mathrm{CMI}(T N_B^{1/\nu};3,1)$.
Therefore, the CMI in the all-region-extensive geometry with temperature
$T$ is equivalent to the CMI of the \emph{three-site} problem with
effective temperature $x = T(N/3)^{1/\nu}=T3^{(n-1)/\nu}$.  
A straightforward calculation (see
App.~\ref{app:maj_three_spin}) then shows that
\begin{equation}
\label{Eq:three-spin-cmi}
\begin{aligned}
\mathrm{CMI}(x;3,1)
&=2h_2\!\left(\frac{2\kappa}{1+3\kappa}\right)
+\frac{1}{1+3\kappa}\ln\frac{1}{1+3\kappa}\\
&\quad
+\frac{3\kappa}{1+3\kappa}
\ln\!\left(\frac{\kappa}{1+3\kappa}\right),
\end{aligned}
\end{equation}  
where $\kappa=e^{-4/x}$, and 
$h_2(u)=-u\ln u-(1-u)\ln(1-u)$ is the binary entropy.  
Deep in the UV ($x\ll1$) and IR ($x\gg1$)  regimes, one finds
\begin{equation}
\mathrm{CMI}(x;3,1)\sim
\begin{cases}
\dfrac{4}{x}e^{-4/x}, & x\ll1,\\[8pt]
\dfrac{1}{2x^2}, & x\gg1 .
\end{cases}
\end{equation}
Fig.~\ref{fig:maj_three_spin_cmi_limits} shows the exact three-spin CMI 
together with these two limiting forms.  In the IR regime, the CMI decays only
algebraically, $\mathrm{CMI}\sim1/(2x^2)$, as $x\to\infty$.  Since
$x=T N_B^{1/\nu}$, at fixed $T>0$ this gives
$\mathrm{CMI}\sim(2T^2)^{-1}N_B^{-2/\nu}$, signaling long-range conditional
correlations.
The UV behavior is even more striking: as $x\to0$, the
CMI has the form $\mathrm{CMI}\sim(4/x)e^{-4/x}$, which decays exponentially fast when $x \to 0$.
This is because in this limit, the
effective three-spin state approaches the equal classical mixture of the
all-up and all-down configurations.  Conditioning on $B$ therefore reveals the shared
orientation between region $A$ and $C$, and thus $I(A:C|B)\to0$. 
 
\subsubsection{Local tripartition.}
For a local tripartition, suppose first that region $A$ and $C$ have length
$b=3^r$ and the buffer $B$ has length $(N'-2)b$, so the total system size is
$N=N'b$.  Then Eq.~\eqref{Eq:maj_UV_independent} gives
\begin{equation}
\mathrm{CMI}(T;N'b,(N'-2)b)
=\mathrm{CMI}(Tb^{1/\nu};N',N'-2).
\end{equation}
Thus it suffices to take $A$ and $C$ to be single-site regions, with $B$
containing the remaining $N-2$ sites.
Setting
$x=T(N/3)^{1/\nu}$, we find that (see App.~\ref{app:maj_local_cmi})
\begin{equation} 
\label{eq:maj_local_cmi_scaling}
\begin{gathered}
\mathrm{CMI}(T;N,N-2) 
=\mathrm{CMI}(x;3,1)
\prod_{r=1}^{n-1}q\!\left(\frac{x}{3^{r/\nu}}\right),\\ 
q(y)=
\left(\frac{2e^{-4/y}}{1+3e^{-4/y}}\right)^2 .
\end{gathered} 
\end{equation} 
Therefore, we have also mapped $\mathrm{CMI}(T;N,N-2) $ to the three-spin problem $\mathrm{CMI}(x;3,1)$, but now with the additional propagating factor $\prod_{r=1}^{n-1}q\!\left({x}/{3^{r/\nu}}\right)$. 
 
Let's first
consider the IR regime $N\gg\xi = T^{-\nu}$.
At fixed $T>0$, the CMI has the algebraic
form
\begin{equation}
\label{eq:maj_local_cmi_high_temperature}
\mathrm{CMI}(T;N,N-2)
\sim \dfrac{R_{\mathrm{CMI}}(T)}{2T^2}\,
(N/3)^{-\log_3 4-2/\nu},
\end{equation}
where $R_{\mathrm{CMI}}(T)=\prod_{k=0}^{\infty}4q(T3^{k/\nu})$ is finite for 
$T>0$ (see App.~\ref{app:maj_local_cmi} for details).
Thus the CMI decays as a power of $N$.  Because the buffer size
$N_B=N-2$ grows with $N$, this power-law decay implies a divergent Markov length.

We next consider the UV regime $N\ll\xi$, i.e., $x=T(N/3)^{1/\nu}\ll1$.
Writing the result in terms of $x$ gives
\begin{equation}
\label{eq:maj_local_cmi_asymptotics}
\begin{aligned}
\mathrm{CMI}(T;N,N-2)
&\sim \dfrac{4}{x}(N/3)^{\log_3 4}\\
&\quad{}\times
\exp\!\left[-\dfrac{8}{x}
\dfrac{(N/3)^{1/\nu}-1}{1-3^{-1/\nu}}-\dfrac{4}{x}\right].
\end{aligned}
\end{equation}
At fixed $x$, this is a power of $N$ multiplied by $\exp[-c(x)N^{1/\nu}]$, with
$c(x)>0$, and therefore decays faster than any
power of $N$.\footnote{The rapid UV decay comes from the product of
$q$ factors in
Eq.~\eqref{eq:maj_local_cmi_scaling}.  Each $q(y)$ is the probability that
information from both side spins is carried through one level of the
hierarchy, which requires each spin to determine the majority of its block.
At low effective temperature, $q(y)\sim4e^{-8/y}$ is exponentially small, so
information from the two microscopic spins is unlikely to reach the top
three-spin problem.}  
We note that this cannot be written as a one-variable scaling form
$\mathrm{CMI}=N^{-\alpha}f(x)$ that one would naively assume.
By contrast, the MI between $A$
and $C$ does have an ordinary scaling collapse (see App.~\ref{app:maj_local_cmi} for details).

\subsubsection{Global tripartition.}  
 
For $N_B=3^r$ and $N\gg N_B$, choosing $b=N_B$ in
Eq.~\eqref{Eq:maj_UV_independent} gives
$\mathrm{CMI}(T;N,N_B)=\mathrm{CMI}(T N_B^{1/\nu};N/N_B,1)$.
Therefore, it suffices to take $B$ to be a single site.  
Taking $N/N_B\to\infty$ at
fixed $x=T N_B^{1/\nu}$ then
defines the scaling function
$F(x)\equiv\lim_{N/N_B\to\infty}\mathrm{CMI}(T;N,N_B)
=\mathrm{CMI}(x;\infty,1)$.  We note that due to the discrete scale of the RG block size $b = 3$, $F(x)$ can oscillate with period one in $\log_{3^{1/\nu}}x$.
 
\begin{figure}[H] 
\centering
\includegraphics[width=\linewidth]{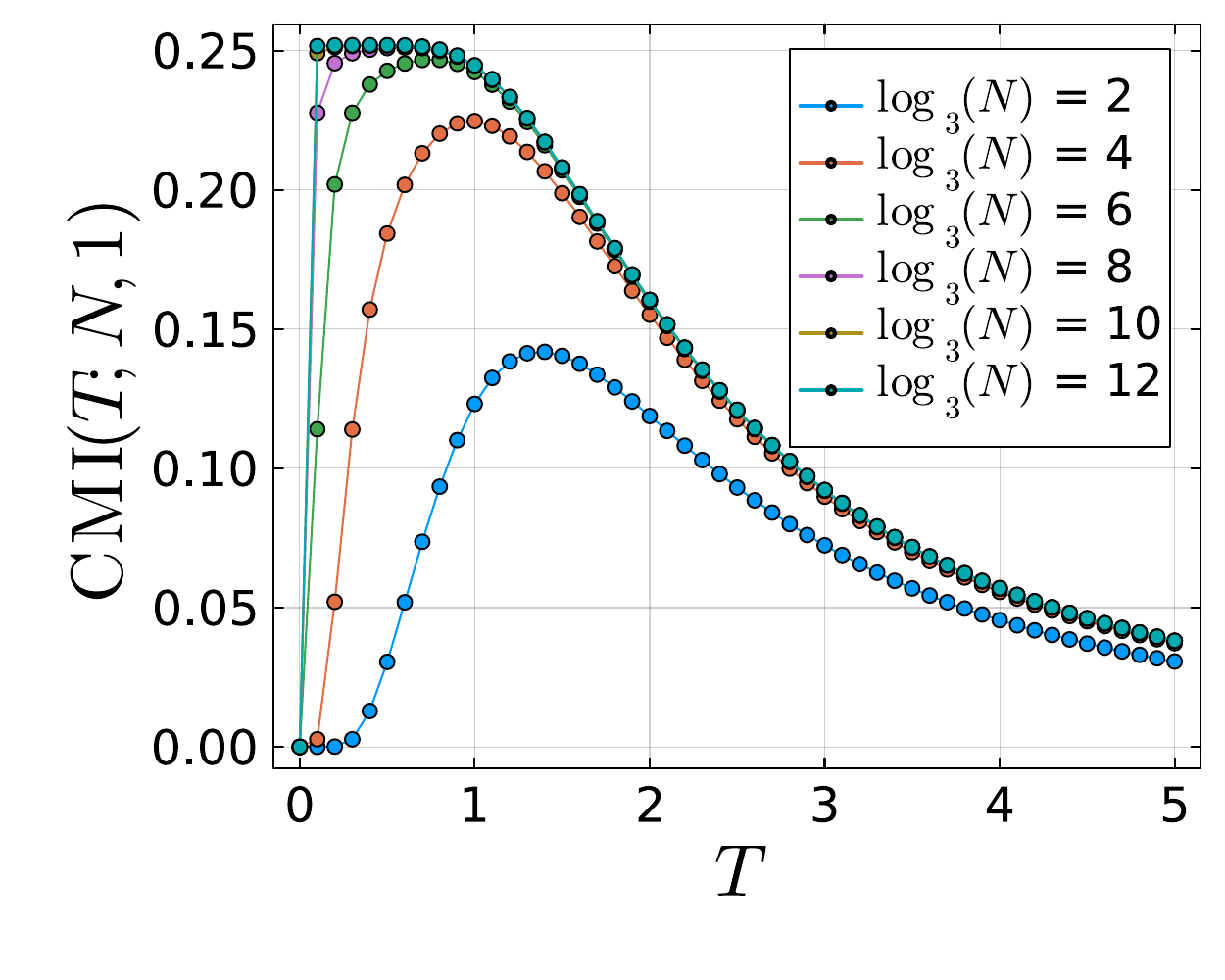} 
\caption{Global-tripartition CMI $\mathrm{CMI}(T;N,1)$ for several $N$ in the 
majority-rule hierarchy at $\nu=2$.  As $N\to\infty$, the curves approach
$F(T)=\mathrm{CMI}(T;\infty,1)$, which obeys
$F(3^{1/\nu}x)\leq F(x)$ by SSA and scales as $O(1)$ for $T\to0$  
}
\label{fig:maj_global_nu2} 
\end{figure}

For the global tripartition, the CMI cannot be reduced to the top three-spin
distribution alone.  We instead evaluate it numerically. 
Fig.~\ref{fig:maj_global_nu2} shows $\mathrm{CMI}(T;N,1)$ as a function of $T$
for several finite system sizes.
For small $N$, we observe that the CMI can first increase and then decrease with $T$.  
However, as $N$
grows, the curves approach the thermodynamic-limit function $F(x)$ with $x=T$ (recall that we set $N_B=1$), which is monotonically decreasing under the discrete sequence $x, 3^{1/\nu} x, \cdots$ due to SSA in Eq.\eqref{eq:global_cmi_scaling_monotonicity}.

Finally, we note that the limits $N\to\infty$ and $T\to0$ do not commute.  For
every finite $N$, the CMI vanishes exactly at $T=0$.  If $N\to\infty$ is taken
first, however, the CMI remains $O(1)$ as $T\to0$, as shown in
Fig.~\ref{fig:maj_global_nu2}.

\section{Stable long-range conditional mutual information on trees}
\label{sec:stable_tree}

Although the majority-rule hierarchical Ising model discussed in
Sec.~\ref{sec:majority_rule_hierarchical_ising} provides a simple
example of the tension between a divergent Markov length and an RG flow toward
a trivial fixed point, it does not exhibit a finite-temperature transition.
Furthermore, both two-point MI and CMI decay polynomially as a function of
distance, and thus both the correlation length and the Markov length diverge.
{
Is there an interesting critical point in one dimension that has zero two-point MI while the two-point CMI decays polynomially?
}
  
It turns out that repeatedly encoding each parent spin into its children along a tree provides a simple framework for constructing such a model~\cite{forney1966concatenated,evans2000broadcasting,yadavalli2025noisy,sommers2025dynamically}.  In fact, the model we will study below can be understood as the classical version of the Bell tree studied in Refs.~\cite{yadavalli2025noisy,sommers2025dynamically}. 
While those works ask whether information about the root spin can be recovered from arbitrarily distant leaves, we instead use RG and CMI to analyze the resulting leaf distribution.

In particular, we find three features that are especially intriguing from the perspective of mixed-state phases.   
(i) At the zero-noise fixed
point, the local CMI vanishes,
while the global CMI remains nonzero.
(ii) The zero-noise fixed point has a strong global (0-form) $\mathbb{Z}_2$
symmetry but is robust against weak local perturbations.  
(iii) At the critical noise strength, the model provides the simplest 
one-dimensional example in which the two-point MI vanishes identically,
whereas the corresponding CMI decays as a power law.

\subsection{SSB and SWSSB fixed points from trees}
\label{sec:two_elementary_fixed}

Let us first motivate the construction using two familiar fixed points: the classical SSB and SWSSB states, which we now briefly review. 
The classical SSB fixed-point state is the equal mixture
\begin{equation}
\label{Eq:SSB}
\rho_{\mathrm{SSB}}
=\frac{1}{2}(|\uparrow\rangle\langle\uparrow|)^{\otimes N}
+\frac{1}{2}(|\downarrow\rangle\langle\downarrow|)^{\otimes N}.
\end{equation}
Throughout, an Ising spin takes values $s=\pm1$, with $s=+1$ 
represented by $|\uparrow\rangle$ and $s=-1$ by $|\downarrow\rangle$ in the Pauli-$Z$ basis.
This state has a weak $\mathbb{Z}_2$ symmetry generated by
$U_X=\prod_j X_j$, namely
$U_X\rho_{\mathrm{SSB}}U_X=\rho_{\mathrm{SSB}}$.
For any tripartition with nonempty $B$, a single spin in $B$ already reveals
which of the two ordered configurations is present.  
Therefore, $I(A:C|B)=0$, and both the local and global Markov
conditions hold (see Tab.~\ref{tab:tree_fixed_point_cmi}). 

Another point that will be useful later is that the state in Eq.~\eqref{Eq:SSB} has a simple tree representation: starting from the root state
$(|\uparrow\rangle\langle\uparrow|+|\downarrow\rangle
\langle\downarrow|)/2$, one can apply the repetition rule
$\uparrow\mapsto\uparrow\uparrow\uparrow$ and
$\downarrow\mapsto\downarrow\downarrow\downarrow$ at every vertex
$\log_3N$ times to generate $\rho_{\mathrm{SSB}}$. 

By contrast, the SWSSB fixed point is the maximally mixed state on the
even-parity subspace,
\begin{equation}
\rho_{\mathrm{SWSSB}}
=\frac{1}{2^{N-1}} 
\sum_{\substack{\mathbf{s}\in\{\pm1\}^N,\\ \prod_j s_j=+1}}
|\mathbf{s}\rangle\langle\mathbf{s}|.
\end{equation}
Unlike the SSB state, $\rho_{\mathrm{SWSSB}}$ is strongly symmetric under
$U_Z=\prod_j Z_j$: because every configuration in its support has even parity,
$U_Z\rho_{\mathrm{SWSSB}}=\rho_{\mathrm{SWSSB}}U_Z
=\rho_{\mathrm{SWSSB}}$.  
For any tripartition into nonempty regions, conditioning on $B$ fixes only the joint parity of $A$ and $C$ and thus leaves one shared
parity bit.  It follows that $I(A:C|B)=\ln 2$, and both Markov conditions fail (see
Tab.~\ref{tab:tree_fixed_point_cmi}).
 
Like the SSB state, the SWSSB state has a simple tree representation: we fix
the root spin to $s=+1$ and apply the parity rule $s\mapsto(s_L,s_R)$, i.e., choosing $(s_L,s_R)$ uniformly subject to $s_L s_R=s$, at every
vertex.

\subsection{Combining SSB and SWSSB rules}

The two tree rules discussed above have complementary properties: the SSB repetition rule suppresses local errors, whereas the SWSSB parity rule stores a global
constraint. 
This allows us to construct states that violate the global Markov condition while satisfying the local one.  
Specifically, we apply the parity rule first and then 
repeating each of its two outputs three times, which corresponds to the stochastic substitution for each parent spin $s$
\begin{equation}
\label{Eq:stable_tree_rule}
|s\rangle\langle s|
\longmapsto
\frac{1}{2}
\sum_{\substack{s_L,s_R=\pm1\\s_Ls_R=s}}
(|s_L\rangle\langle s_L|)^{\otimes3}
\otimes(|s_R\rangle\langle s_R|)^{\otimes3}.
\end{equation}
Starting from a fixed root $s=+1$ and applying Eq.~\eqref{Eq:stable_tree_rule}
for $n$ steps produces the classical mixed state $\rho$ on $N=6^n$ leaves.  
We call this construction
the \textit{SSB--SWSSB tree}, which is the classical version of the
Bell-tree model studied in
Refs.~\cite{yadavalli2025noisy,sommers2025dynamically}. 

We remark that $\rho$ has both a \emph{strong} $\mathbb{Z}_2$ symmetry,
generated by $U_Z=\prod_j Z_j$, and a weak $\mathbb{Z}_2$ symmetry,
generated by $U_X=\prod_j X_j$.
The strong symmetry follows because each substitution preserves parity:
$s_L^3 s_R^3=s_L s_R=s$.
Because the root is fixed to $+1$, every configuration in
$\rho$ has even parity. Hence $U_Z\rho=\rho U_Z=\rho$.
By contrast, the weak symmetry follows because $\prod_j X_j$ merely exchanges the two
equiprobable terms in Eq.~\eqref{Eq:stable_tree_rule}, so
$U_X\rho U_X=\rho$.

Let us now show that $\rho$ violates the global Markov condition while satisfying the local one.
Because of the repetition rule, the value of any single spin can always be inferred from one of its neighbors. Thus, the CMI vanishes when
$N_A=N_C=O(1)$.
By contrast, the parity rule propagates information about the parent to both sides. A simple counting argument in
App.~\ref{app:stable_tree_bmax_counting} shows that the CMI is $\log 2$ whenever
$N_B\le B_{\text{max}}$ (and vanishes otherwise), where
\begin{equation}
\label{eq:stable_tree_bmax}
B_{\text{max}}=\frac{N-2b}{2b-1} = \frac{N-6}{5}.
\end{equation}
Here $b$ is the size of one repetition block, which is $3$ in our model. 
Since $B_{\text{max}}$ is
extensive in $N$, any subextensive buffer, i.e., $N_B(N)/N\to0$, eventually
yields a CMI of $\log 2$.
Table~\ref{tab:tree_fixed_point_cmi} summarizes the local and global CMI of the
SSB, SWSSB, and SSB--SWSSB fixed points.
\begin{table}[t]
\centering
\renewcommand{\arraystretch}{1.5}
\begin{tabular}{lcc}
\hline\hline
Fixed point & Local ($\frac{N_A}{N_B}\to0$) & Global ($\frac{N_A}{N_B}\to\infty$) \\
\hline
SSB                    & $0$      & $0$      \\
SWSSB                  & $\ln 2$  & $\ln 2$  \\
SSB--SWSSB tree        & $0$      & $\ln 2$  \\
\hline\hline  
\end{tabular}
\caption{
The local and global CMI of the
SSB, SWSSB, and SSB--SWSSB fixed points. 
More generally, for the SSB--SWSSB tree in the thermodynamic limit,
$I(A:C|B)=0$ when $N_A/N_B<2$ and $I(A:C|B)=\ln 2$ when $N_A/N_B\ge 2$.
}
\label{tab:tree_fixed_point_cmi}
\end{table}

\subsection{Ideal RG} 
\label{sec:tree_ideal_RG}
  
We now apply independent spin-flip noise of strength $p$ to the
SSB--SWSSB tree and study the corresponding probability using RG.
The most natural and intuitive RG scheme is to reverse the
two steps of Eq.~\eqref{Eq:stable_tree_rule}: take a majority vote within
each repetition triple and then multiply the two majority signs.  
As shown in
App.~\ref{app:tree_nonideal_rg}, this map closes on a single effective error 
probability and identifies an unstable (critical) fixed point at
$p_c^{\mathrm{nonideal}}\simeq0.225$.  Despite its intuitive form and useful
properties, this RG scheme is \textit{nonideal} because it does not preserve
correlations.

Let us derive an ideal RG transformation based on the correlation-preserving criterion.
The
essential step is to exploit the special tree structure.
Let
$S$ denote the observed spins in a block, $Y$ all spins outside the block,
and $\sigma$ the hidden parent spin.
The block is connected to the rest of the tree
only through $\sigma$, so $P(S,Y|\sigma)=P(S|\sigma)P(Y|\sigma)$.
Therefore, all information that $S$ contains about $Y$ is encoded in the posterior probability of $\sigma$ given $S$:
\begin{equation}
q(S)=P(\sigma=+1|S).
\end{equation}
In particular, $P(Y|S)=q(S)P(Y|\sigma=+1)+[1-q(S)]P(Y|\sigma=-1)$, which implies
\begin{equation}
I(S:Y)-I(q(S):Y)=I(S:Y|q(S))=0.
\end{equation}
Keeping $q(S)$ therefore satisfies the correlation-preserving condition in
Eq.~\eqref{Eq:correlation_preserving}. 
We derive the recursion for $q(S)$ in App.~\ref{app:ideal_tree_rg}. 

Since $q(S)$ is continuous on $[0,1]$ and can therefore take infinitely many
values, in what sense does retaining $q(S)$ constitute an ideal RG?  We first
note that, by definition, $q(S)$ is a function of $S$ and thus cannot take more
values than there are configurations of $S$.  As the system size increases,
more and more distinct configurations of $S$ share the same value of $q(S)$.
Therefore, $q(S)$ indeed coarse-grains the configuration space of $S$.  
However, because $q(S)$ is continuous, this ideal RG does not retain an
effective variable of fixed finite dimension, and thus the UV-finiteness argument
of Sec.~\ref{sec:UV_finite} does not apply directly.  Nevertheless,
we show analytically below that the CMI of the noisy SSB--SWSSB tree is still
UV-finite. In fact, it is bounded above by $\log 2$.

\subsection{Conditional mutual information}
\label{sec:tree_cmi}

We now study the CMI using the posterior probabilities $q(S)$.  We take the regions $A$ and $C$ to have equal size, $N_A=N_C$, and write the CMI as
$\mathrm{CMI}(p;N,N_B)$, where $N=6^n=N_A+N_B+N_C$.  The root spin is fixed to
$\Sigma=+1$.

Before doing any actual calculations, let's first show the UV-finiteness of CMI. Since $\mathrm{CMI}(p;N,N_B)$ is a monotonically decreasing function of $N_B$ with $N$ fixed, it suffices to show that  $\lim_{N\rightarrow\infty}\mathrm{CMI}(p;N,N_B)$ for a fixed $N_B$ is bounded.

Recall the SSB--SWSSB generation rule in Eq.~\eqref{Eq:stable_tree_rule}, i.e., $|s\rangle\langle s|
\mapsto
\frac{1}{2}
\sum_{\substack{s_L,s_R=\pm1,\ s_Ls_R=s}}
(|s_L\rangle\langle s_L|)^{\otimes3}
\otimes(|s_R\rangle\langle s_R|)^{\otimes3}$. 
The fixed root $\Sigma=+1$ therefore produces either
$s_L=s_R=+1$ or $s_L=s_R=-1$, each with probability $1/2$. 
Denoting their common value as
$\eta$, one then has the entropy $H(\eta)=\log 2$. 
Crucially, once $\eta$ is known, $A$ and $C$ are independent, which also implies $I(A:C|B,\eta) = 0$. It follows that 
\begin{equation}
  \label{eq:stable_tree_global_binary_bound}
\begin{aligned}
I(A:C|B) & \leq I(A:C\eta|B) = I(A:\eta|B) + I(A:C|B \eta)\\
& = I(A:\eta|B) \leq H(\eta) = \log 2,
\end{aligned}
\end{equation}
where the first inequality follows from SSA, and the second inequality uses the fact that $I(A:\eta|B) = H(\eta|B) - H(\eta|AB) \leq H(\eta|B) \leq H(\eta)$ (since $\eta$ is a classical bit, its conditional entropy is positive).
We note that Eq.~\eqref{eq:stable_tree_global_binary_bound} holds at every depth and noise strength.
The CMI is therefore UV-finite even though the ideal RG does not retain a finite Hilbert space dimension.

Below, we compute CMI for the following three geometries by standard population dynamics~\cite{mezard2009information}. 
(i) All-region-extensive: all three regions grow with $N$, and
$N_B=B_{\text{max}} = (N-6)/5$.  (ii) Local:
$N_B=N-2$, so $A$ and $C$ are single sites at the two endpoints.  (iii) Global:
$N_B$ is any fixed even integer, and we take $N\rightarrow\infty$ at fixed
$N_B$.

\begin{figure*}[t]
\centering
\includegraphics[width=\linewidth]{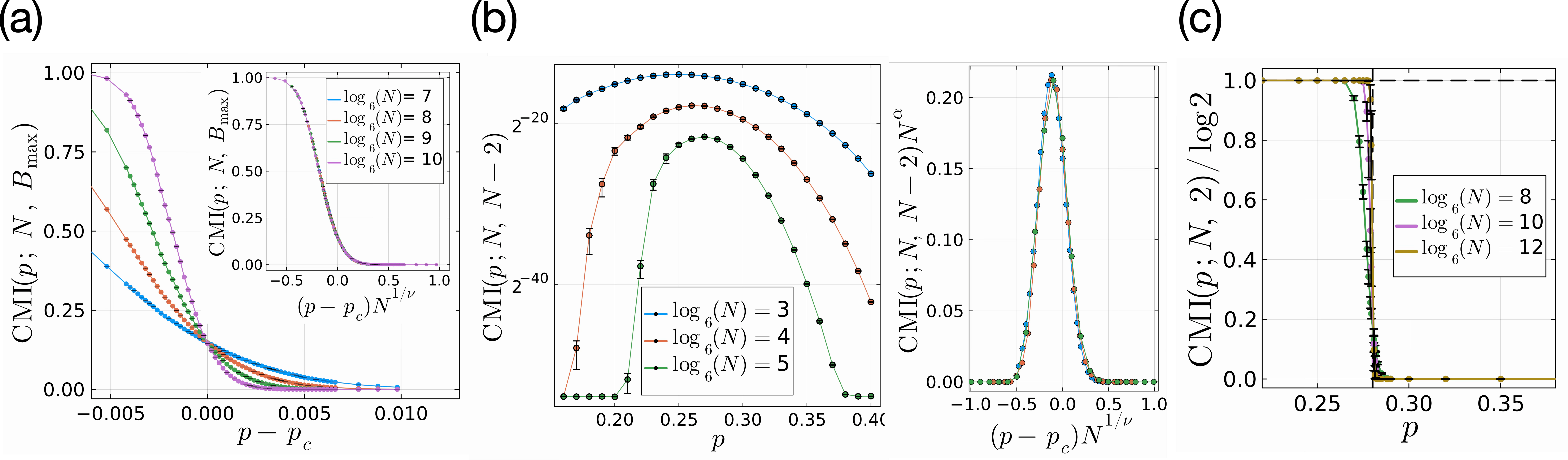}
\caption{CMI of SSB-SWSSB tree.  (a)
All-region-extensive CMI for $N_B=B_{\mathrm{max}}$ near the transition.  The
inset shows the same data against $(p-p_c)N^{1/\nu}$, using
$p_c\simeq0.2802$ and $\nu\simeq3.9$.  (b) Local CMI for $N_A=N_C=1$ and
$N_B=N-2$.  The left panel shows $\mathrm{CMI}(p;N,N-2)$ for
$N=6^3,6^4,6^5$, and the right panel shows the data collapse obtained by
plotting $N^\alpha\mathrm{CMI}(p;N,N-2)$ against $(p-p_c)N^{1/\nu}$, using
$p_c=0.2802$, $\nu=3.9$, and $\alpha=1.5$.  (c) Global CMI at fixed $N_B=2$
for $N=6^8,6^{10},6^{12}$, normalized by $\log 2$.  The horizontal dashed
line marks the predicted low-noise value, and the vertical dashed line marks
$p_c=0.2802$.  
}
\label{fig:concat_all}
\end{figure*}

\subsubsection{All-region-extensive geometry}  

Fig.~\ref{fig:concat_all}(a) shows the CMI as a function of $p$ for various $N$, with $N_B=B_{\text{max}} = (N-6)/5$.  
The CMI curves clearly cross near $p_c\simeq0.2802$ for all $N$.  
Furthermore, data for different system sizes collapse nicely on one curve when plotted against
$(p-p_c)N^{1/\nu}$, with $\nu\simeq3.9$, as shown in the inset.  
This provides strong evidence that the CMI in this tripartition can be described by a universal scaling function $f((p-p_c)N^{1/\nu})$.
The estimates of $p_c$ and $\nu$ obtained here will be used below to compute the CMI for local and global tripartitions.

\subsubsection{Local tripartition}
\label{sec:tree_cmi_local}

We next choose $N_A=N_C=1$ and $N_B=N-2$, so $A$ and $C$ are the first and
last leaves of the depth-$n$ tree.
The left panel of Fig.~\ref{fig:concat_all}(b) shows
$\mathrm{CMI}(p;N,N-2)$ as a function of $p$ for different $N$.
As $N$ 
grows, the peak near $p_c$ becomes narrower. 
Furthermore, using the values of $p_c$ and
$\nu$ found above, we find that the data for different $N$ are described by
\begin{equation}
\mathrm{CMI}(p;N,N-2)
=N^{-\alpha}f\!\left((p-p_c)N^{1/\nu}\right),
\end{equation}
where $\alpha \simeq 1.5$ [see Fig.~\ref{fig:concat_all}(b)].
{
This suggests a power-law decaying CMI at $p=p_c$, , i.e.,
$\mathrm{CMI}(p_c;N,N-2)\sim N^{-\alpha}$.  
On the other hand, away from $p_c$, the CMI decreases much more rapidly over the accessible system sizes. 
}
 
We close by contrasting the CMI with the two-point mutual information.  
For classical Ising variables, a vanishing connected two-point correlation $\langle Z_i Z_j\rangle_c = \langle Z_i Z_j \rangle - \langle Z_i \rangle \langle Z_j \rangle$ implies a vanishing MI.
When $p = 0$, the initial state has a weak $\mathbb{Z}_2$ symmetry generated by $\prod_j X_j$, and thus $\langle Z_i\rangle = 0$ for all $i$.  Furthermore, the parity rule gives $\langle Z_i Z_j\rangle = 0$ whenever $|i-j| \geq 3$ and $\log_6(N)\geq 2$. 
Therefore, the connected two-point correlation function vanishes, i.e., $\langle Z_i Z_j\rangle_c = 0$.
Since independent bit-flip noise acts locally and cannot correlate initially independent spins, two-point MI remains zero for all $p$.
This transition contrasts with equilibrium criticality, where the
mutual information decays algebraically and the CMI is short-ranged. 
 
\subsubsection{Global tripartition} 
Finally, we consider the global tripartition
$N_A=N_C=(N-N_B)/2$ for any fixed even $N_B$.  This tripartition can be understood as first taking
$N=6^n\rightarrow\infty$ at fixed $N_B$ and then studying
$\mathrm{CMI}(p;\infty,N_B)
:=\lim_{n\rightarrow\infty}\mathrm{CMI}(p;6^n,N_B)$.

Before doing any explicit calculation, we first demonstrate the power of SSA when combined with the CMI scaling assumption.
Recall that SSA implies $\mathrm{CMI}(p;\infty, 6N_B)\le \mathrm{CMI}(p;\infty,N_B)$.
Combining this with the scaling assumption $\mathrm{CMI}(p;\infty, N_B)\sim F(x),
\ x=(p-p_c)N_B^{1/\nu}$, one finds (we take $N_B=N_{B,0}6^m$)
\begin{equation}
  \label{eq:stable_tree_global_scaling_ssa}
F\!\left(6^{1/\nu}x\right)\le F(x).
\end{equation}
For $x<0$, a repeated use of Eq.\eqref
{eq:stable_tree_global_scaling_ssa} implies $F(x)\ge F(-\infty)=\log 2$, where the last equality follows because the low-noise phase flows to the SSB-SWSSB
fixed point. 
Furthermore, since the CMI is bounded above by $\log 2 $ as we've showned in Eq.~\eqref{eq:stable_tree_global_binary_bound}, one finds
\begin{equation}
  \label{eq:stable_tree_global_low_scaling}
F(x <0)=\log 2.
\end{equation}
Therefore, we anticipate that the CMI in the global tripartition with $p<p_c$ is always $\log2$.
On the other hand, for $x>0$, SSA gives only the lower bound $F(x)\ge0$ and does
not determine the value.  We therefore turn to the numerics.

Fig.~\ref{fig:concat_all}(c) shows $\mathrm{CMI}(p;N,2)$ for increasing $N$.  We find that when $p<p_c$, the data approach the plateau at $\log 2$, verifying the prediction in
Eq.~\eqref{eq:stable_tree_global_low_scaling}. 
On the other hand, when $p>p_c$, where SSA is not constraining at all, the data surprisingly approach almost exactly zero as $N$ grows.  The numerics therefore suggest
that, for every fixed $N_B$, 
\begin{equation}
\mathrm{CMI}(p;\infty,N_B)
=
\begin{cases}
\log 2, & p<p_c,\\
0, & p>p_c.
\end{cases}
\label{eq:stable_tree_global_phases}
\end{equation}

We now provide an argument on why the CMI vanishes as long as $p>p_c$, where SSA and the scaling assumption together are not constraining, using the additional structure of the SSB-SWSSB tree. 
When $p>p_c$, one expects the system to flow to the high-noise phase, in which no
information about $\eta$ (the common value in the first generation of the fixed root $\Sigma = 1$) survives in the leaves. 
This implies that the posterior $P(\eta=+1|ABC)$
concentrates at $1/2$, and thus
$I(\eta:ABC)\to0$.
Now, the chain rule gives 
$I(\eta:ABC)=I(\eta:B)+I(\eta:A|B)+I(\eta:C|AB) \ge I(A:\eta|B)$.  Combining with the intermediate quantity of Eq.~\eqref{eq:stable_tree_global_binary_bound}, one finds
\begin{equation}
0\le I(A:C|B)\le I(A:\eta|B)\le I(\eta:ABC)\to0 .
\end{equation}
Consequently, $\mathrm{CMI}(p;\infty,N_B)=0$ for every $p>p_c$, and $F(x)=0$ for 
$x>0$.
This accounts for the second line of Eq.~\eqref{eq:stable_tree_global_phases}, and together with
Eq.~\eqref{eq:stable_tree_global_low_scaling} it explains both plateaus seen in
Fig.~\ref{fig:concat_all}(c).

\section{Conclusion and Outlook}
\label{sec:conclusion}

In this work, we studied how the renormalization group (RG) constrains the conditional mutual information (CMI), and then used the RG to
analyze hierarchical models with long-range CMI. 
Our starting point was the ideal (locally reversible) RG scheme defined in
Ref.~\cite{sang2024mixed}.
We showed that the mere \emph{existence} of such a scheme with a fixed on-site
Hilbert space dimension forces the CMI to be UV-finite.
Assuming the scaling limit exists, this result implies that the CMI is
\textit{invariant} under RG when all three regions are rescaled together.
Although RG is expected to discard short-distance information while preserving
long-distance properties, it is often unclear what qualifies as a genuine
long-distance property. Our result identifies the CMI as one such property. We
discuss this point further below, along with cases in which the CMI is not
UV-finite.

We then study two hierarchical models whose RG can be carried out exactly.
The first model is a majority-rule variant of Dyson's hierarchical Ising model.
We find that it has a divergent Markov length at every finite temperature,
$0<T<\infty$, yet an ideal RG drives the model to the infinite-temperature
product state.
The state is connected to this product state in both directions by circuits
with sublinear range and thus satisfies the two-way connection criterion of
Ref.~\cite{sang2024mixed}, but it is not phase-equivalent to the product state
under the polylogarithmic-range  definition of Ref.~\cite{sang2025mixed}.    
The second model is a tree formed by concatenating the rules that generate the
classical SSB and SWSSB fixed points, which is the classical counterpart of the
Bell-tree model studied in
Refs.~\cite{yadavalli2025noisy,sommers2025dynamically}.
Notably, it respects the local Markov property while violating the global one.
Under independent spin-flip noise, the local CMI decays polynomially as a function of $N$ at the critical noise strength, while the
two-point mutual information vanishes identically.
 
Several directions naturally emerge from our results.
First, the argument of Sec.~\ref{sec:UV_finite} is not specific to the CMI, and in fact it can be directly generalized to the mutual information
between two \emph{separated} regions. 
This is because the argument uses only the correlation-preserving condition and the fact that enlarging $A$ and $C$ cannot decrease the CMI, both of which also hold for the MI.  
It would be interesting to look for other quantities that are guaranteed to be UV-finite due to the existence of an ideal RG.  Like the CMI, we anticipate that these quantities are invariant under RG in the scaling limit and thus qualify as genuine long-distance properties.

Second, there exist examples where the CMI is not UV-finite, and the most familiar one is a Fermi surface in $d\ge2$, whose entanglement entropy
violates the boundary law with scaling
$(L/a)^{d-1}\log(L/a)$~\cite{wolf2006violation,gioev2006entanglement,
swingle2010entanglement}.
Consistent with our criterion, the RG scheme that reproduces the entropy
scaling is the branching MERA~\cite{evenbly2014class,evenbly2014scaling,swingle2016constructions}, in
which the coarse-grained state splits into an increasing number of decoupled
copies at every step, so that the on-site dimension grows with the RG scale. 
Another example is that of fracton
systems~\cite{chamon2005glassiness,Haah2011local,vijay2016fracton},
whose entanglement
renormalization also bifurcates into several
copies~\cite{haah2014bifurcation,dua2020bifurcating}.
We note that Ref.~\cite{swingle2016} extends the $s$-source classification in Ref.~\cite{swingle2016constructions} to
mixed states, with slightly different definitions of correlation preservation.
It would be interesting to construct a mixed-state branching RG that preserves
correlations in the sense of Eq.~\eqref{Eq:correlation_preserving}, and then to
identify which quantities are UV-finite.

Third, the two hierarchical models discussed in Secs.~\ref{sec:majority_rule_hierarchical_ising} and~\ref{sec:stable_tree} provide interesting examples for testing learnability ~\cite{barratt2022transitions,singh2025mixed,chen2025learning,feng2025hardness,hou2025machine,schuster2025hardness,kumar2026unlearnable}, i.e., whether one can learn the state from a number of measurement outcomes that grows at most polynomially with the system size. 
In particular, Ref.~\cite{kumar2026unlearnable} recently showed that a finite Markov length is sufficient for efficiently learning a quantum state, whereas a state with constant CMI in the local tripartition is unlearnable under certain assumptions.
It is therefore interesting to test probability distributions in the regime not settled by Ref.~\cite{kumar2026unlearnable}, namely, those with infinite Markov length whose CMI decays polynomially rather than saturating to a constant.
From the RG perspective, one would anticipate that the majority-rule hierarchical model is easier to learn than the critical point of the SSB--SWSSB tree, since the former flows to a trivial state while the latter is a genuinely nonequilibrium critical point.
However, the answer may also depend on the learning scheme.  Decoherence of the SSB--SWSSB fixed point provides a natural diffusion-model
path~\cite{hu2025local,liu2025measurement}, so if the fixed point is learnable by a diffusion model, its critical point is also learnable.
It would be interesting to settle these questions. 

\acknowledgments 
I am especially grateful to Tarun Grover for many inspiring discussions and for encouraging me to write this paper.
I also thank John McGreevy, Tim Hsieh, Xiang Li, Kaixiang Su, and Yi-Zhuang You for helpful discussions.   
I acknowledge the use of generative AI tools (Claude and Codex) to assist with writing numerical code, generating figures, proofreading the manuscript, and discussing the work. 
This research was supported in part by the following grants to the Kavli Institute for Theoretical Physics (KITP): NSF PHY-2309135, Heising-Simons Foundation, and the Simons Foundation International  SFI-MPS-T-Institutes-00021142, LB.

\bibliography{bibs}
 
\clearpage
\appendix
  
\section{Correlation preservation in the partial circuit}
\label{app:modified_circuit_recovery}

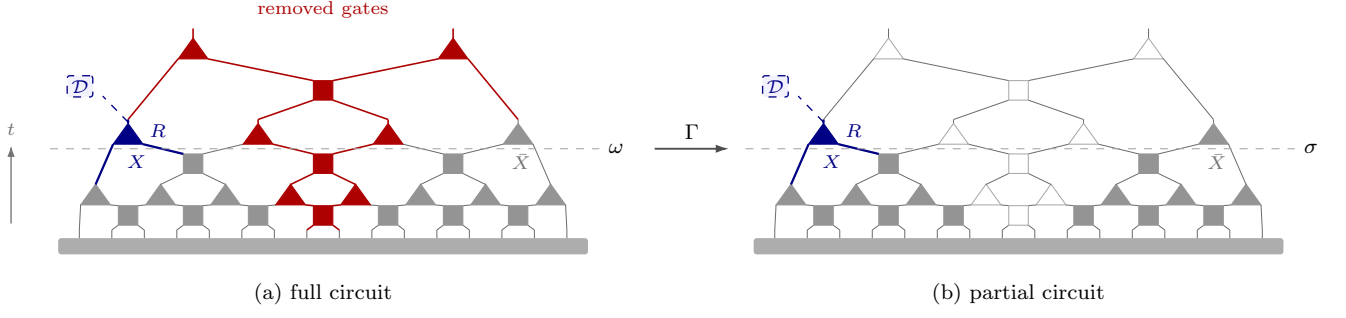
\begin{figure*}[!t]
\centering
\begin{tikzpicture}[
  x=0.43cm, y=0.43cm, font=\footnotesize,
  wire/.style={draw=black!58, line width=0.35pt},
  redwire/.style={draw=red!68!black, line width=0.6pt},
  bluewire/.style={draw=blue!52!black, line width=0.85pt},
  gategray/.style={fill=black!42, draw=black!42, line width=0.2pt},
  gatered/.style={fill=red!68!black, draw=red!68!black, line width=0.2pt},
  gateghost/.style={fill=white, draw=black!35, line width=0.3pt},
  gateR/.style={fill=blue!52!black, draw=blue!52!black, line width=0.2pt},
  lab/.style={font=\scriptsize, fill=white, inner sep=0.8pt},
  gam/.style={-{Latex[length=1.8mm,width=1.1mm]}, draw=black!70, line width=0.7pt}
]
\newcommand{\dg}[6]{%
  \draw[#6] ({#1-#2/2},#3) -- ({#1-0.3},{#3+0.32*#4-0.3});
  \draw[#6] ({#1+#2/2},#3) -- ({#1+0.3},{#3+0.32*#4-0.3});
  \draw[#6] ({#1-0.3},{#3+0.32*#4+0.3}) -- ({#1-#2+0.46},{#3+0.78*#4-0.32});
  \draw[#6] ({#1+0.3},{#3+0.32*#4+0.3}) -- ({#1+#2-0.46},{#3+0.78*#4-0.32});
  \draw[#5] ({#1-0.3},{#3+0.32*#4-0.3}) rectangle ({#1+0.3},{#3+0.32*#4+0.3});}
\newcommand{\ig}[6]{%
  \draw[#6] (#1,{#3+0.78*#4+0.32}) -- (#1,{#3+#4});
  \draw[#5] ({#1-0.46},{#3+0.78*#4-0.32}) -- ({#1+0.46},{#3+0.78*#4-0.32})
            -- (#1,{#3+0.78*#4+0.32}) -- cycle;}
\newcommand{\eg}[5]{%
  \draw[#5] (#1,#3) -- ({#2+(#1>#2 ? 0.46 : -0.46)},{#3+0.78*#4-0.32});}
\newcommand{\mera}[2]{%
  \fill[black!32, rounded corners=1.2pt] (-0.65,-0.75) rectangle (15.65,-0.25);
  \foreach \i in {0,...,15} {\draw[wire] (\i,-0.25) -- (\i,0);}
  \foreach \c in {1.5,3.5,5.5,9.5,11.5,13.5} {\dg{\c}{1}{0}{1.4}{gategray}{wire}}
  \foreach \m in {0.5,2.5,4.5,10.5,12.5,14.5} {\ig{\m}{1}{0}{1.4}{gategray}{wire}}
  \dg{7.5}{1}{0}{1.4}{#1}{#2}
  \foreach \m in {6.5,8.5} {\ig{\m}{1}{0}{1.4}{#1}{#2}}
  \eg{0}{0.5}{0}{1.4}{wire} \eg{15}{14.5}{0}{1.4}{wire}
  \foreach \c in {3.5,11.5} {\dg{\c}{2}{1.4}{2}{gategray}{wire}}
  \ig{13.5}{2}{1.4}{2}{gategray}{wire}
  \dg{7.5}{2}{1.4}{2}{#1}{#2}
  \foreach \m in {5.5,9.5} {\ig{\m}{2}{1.4}{2}{#1}{#2}}
  \eg{14.5}{13.5}{1.4}{2}{wire}
  \dg{7.5}{4}{3.4}{2.8}{#1}{#2}
  \foreach \m in {3.5,11.5} {\ig{\m}{4}{3.4}{2.8}{#1}{#2}}
  \eg{1.5}{3.5}{3.4}{2.8}{#2} \eg{13.5}{11.5}{3.4}{2.8}{#2}
  \draw[bluewire] (3.2,2.34) -- (1.96,2.64);
  \eg{0.5}{1.5}{1.4}{2}{bluewire}
  \ig{1.5}{2}{1.4}{2}{gateR}{bluewire}
  \draw[blue!52!black, dashed, line width=0.5pt] (1.5,3.32) -- (0.72,4.12);
  \node[draw=blue!52!black, dashed, line width=0.5pt, fill=blue!5,
        rounded corners=1pt, inner sep=1.6pt, font=\scriptsize,
        text=blue!52!black] at (0.02,4.42) {$\mathcal D$};
  \node[font=\scriptsize, text=blue!52!black] at (2.42,3.05) {$R$};
  \draw[dashed, gray!70] (-0.9,2.5) -- (16.0,2.5);
  \node[lab, text=blue!52!black] at (1.76,2.10) {$X$};
  \node[lab, text=black!55] at (13.6,2.10) {$\bar X$};}

\begin{scope}
  \mera{gatered}{redwire}
  \node[anchor=west, font=\footnotesize] at (16.0,2.5) {$\omega$};
  \node[font=\scriptsize, text=red!68!black] at (7.5,6.8) {removed gates};
  \draw[-{Latex[length=1.5mm,width=1mm]}, black!55, line width=0.5pt]
    (-2.1,0.2) -- (-2.1,2.6);
  \node[font=\scriptsize, text=black!55] at (-2.1,3.1) {$t$};
  \node[font=\footnotesize] at (7.5,-1.95) {(a) full circuit};
\end{scope}
\begin{scope}[xshift=9.2cm]
  \mera{gateghost}{wire}
  \node[anchor=west, font=\footnotesize] at (16.0,2.5) {$\sigma$};
  \node[font=\footnotesize] at (7.5,-1.95) {(b) partial circuit};
\end{scope}
\draw[gam] (17.7,2.5) -- (20.0,2.5);
\node[font=\footnotesize, anchor=south, inner sep=2pt] at (18.85,2.62) {$\Gamma$};
\end{tikzpicture}
\caption{
Schematic illustration of correlation preservation in the partial circuit.
Time runs upward and the state at the bottom is the initial state $\rho$. 
The red gates in (a) form the causal cone removed in the partial circuit (b).  
The blue triangle represents a retained gate $R$ whose support $X$ consists of two sites in this plot. 
Undoing the removed gates one at a time, starting from the latest, turns the
input $\omega$ of $R$ in the full circuit into
$\sigma=(\operatorname{id}_X\otimes\Gamma)(\omega)$, with $\Gamma$ supported on
$\bar X$. By Eq.~\eqref{eq:recovery_stability}, the same recovery channel
$\mathcal D$ (dashed box) undoes $R$ in both circuits.
}
\label{fig:partial_circuit_recovery}
\end{figure*}

Recall that steps 1 and 3 of Sec.~\ref{sec:UV_finite} apply only part of the ideal RG circuit. 
Since correlation preservation is a property of both the gate and the state, we need to show that the partial RG circuit is still correlation preserving. 
We formulate this precisely by the following lemma.

\begin{lemma}
Consider the full RG circuit, whose gates are applied in causal order and are
each correlation preserving on the input they receive there. 
Construct a partial circuit by removing a subset of gates such that, whenever
a gate is removed, every later gate whose input can depend on its output is
also removed.
Then every retained gate is
correlation preserving on the input it receives in the partial circuit.
\end{lemma}

Let us first note a simple fact: exact recovery is insensitive to what happens outside the
support of a gate. 
Specifically, suppose a gate $R$ supported on $X$ is exactly recovered by $\mathcal D$ on a
state $\omega$, and let $\sigma=(\operatorname{id}_X\otimes\Gamma)(\omega)$, where $\Gamma$ is any channel acting only on $\bar X$.  Since
$\Gamma$ commutes with both $R$ and $\mathcal D$, 
\begin{equation}
\left[(\mathcal D\circ R)\otimes\operatorname{id}\right](\sigma)
= 
(\operatorname{id}_X\otimes\Gamma)
\left[(\mathcal D\circ R)\otimes\operatorname{id}\right](\omega)
=\sigma,
\label{eq:recovery_stability}
\end{equation}
so the same $\mathcal D$ still recovers $R$ on $\sigma$.
.

We next show that the two inputs of a retained gate differ by exaclty such an operation.
Fig.~\ref{fig:partial_circuit_recovery} summarizes the argument on a
MERA-like circuit. 
Let $R$ be a retained gate with support $X$, and let $\omega$ and $\sigma$ be
the inputs it receives in the full and in the partial circuit.  
By construction, the two differ
only through the removed gates that precede $R$, and each of these acts outside
$X$.
We now undo those gates from $\omega$ one at a time, starting from the latest.
Let $S$ be the latest removed gate still present at a given stage, and let
$\mathcal E$ be its recovery channel, which exists because $S$ is correlation
preserving on its own input in the full circuit.  Every gate applied after $S$
is retained, so by the removal rule it acts outside the support of $S$ and
commutes with $\mathcal E$.  Applying $\mathcal E$ therefore undoes $S$ exactly
and produces the state of the circuit in which $S$ is absent.
Iterating until no removed gate is left turns $\omega$ into $\sigma$, so that
\begin{equation}
\sigma
=
(\operatorname{id}_{X}\otimes\Gamma)(\omega),
\label {eq:hybrid_input_from_full}
\end{equation}
where $\Gamma$ is the composite of these recoveries.  Each of them is supported
on a removed gate and hence acts outside $X$, so
Eq.~\eqref{eq:recovery_stability} applies and shows that the recovery of $R$ on
$\omega$ remains valid on $\sigma$, which proves the lemma.

\section{Details for the majority-rule hierarchical Ising model}
\label{app:maj_hierarchical_details}

\subsection{Derivation of Eq.~\eqref{eq:maj_rel_entropy}}
\label{app_sec:rel}
In this appendix we derive the relative entropy
of the renormalized state $\rho_k=\rho(T_k,N_k)$ from the maximally mixed state.
We write $N=3^n$, and recall that $N_k=N/3^k$ and $T_k=T3^{k/\nu}$ are the system
size and temperature after $k$ RG steps.
We first note that the relative entropy is simply $D\!\left(\rho_k\middle\|{\mathbbm{1}}/{2^{N_k}}\right)
=N_k\log2-S(\rho_k)$, which follows directly from
$\log(\mathbbm{1}/2^{N_k})=-(N_k\log2 )\mathbbm{1}$.
It therefore suffices to compute $S(\rho_k)$, which for a Gibbs state is
$S(\rho_k)=\log Z_k-\beta_k\partial_{\beta_k}\log Z_k$ with $\beta_k = 1/T_k$.
Since $\beta_k$ decreases along the RG flow, we can evaluate this expression using a high-temperature expansion.

Repeated use of the fixed-majority block sum in
Eq.~\eqref{eq:app_block_weight} factorizes the partition function as
\begin{equation}
Z_k=2\prod_{p=0}^{n-k-1} 
\left(e^{ 3^{1-p/\nu}\beta_k}+3e^{- 3^{-p/\nu}\beta_k}\right)^{{N_k}/3^{p+1}}.
\end{equation}
Since $\log(e^{3x}+3e^{-x})=\log4+3x^2/2+O(x^3)$, summing over the levels gives
$\log Z_k=N_k\log2+ \beta^2_k\langle H_{N_k}^2\rangle_0/2+O({N_k}\beta_k^3)$,
where $\langle\cdot\rangle_0$ is the average in the maximally mixed state and,
because every coupling squares to unity while distinct couplings are orthogonal
in that state,
\begin{equation}
\begin{aligned}
\left\langle H_{N_k}^2\right\rangle_0
&=N_k\sum_{p=0}^{n-k-1}3^{-p(1+2/\nu)}\approx A_{\nu}N_k,
\end{aligned}
\end{equation}
where $A_\nu={1}/{(1-3^{-(1+2/\nu)})}$.
The identity $D=\beta_k\partial_{\beta_k}\log Z_k-\log Z_k+N_k\log2$ then gives
\begin{equation}
\begin{split}
D \Big(\rho_k ||\frac{\mathbbm{1}}{2^{N_k}}\Big)
&\approx \frac{A_{\nu}N_k}{2T_k^{2}}=\frac{A_\nu}{2}\,\frac{N}{T^{2}3^{k(1+2/\nu)}},
\end{split}
\end{equation}
which is the desired result.

\subsection{Derivation of Eq.~\eqref{Eq:three-spin-cmi}} 
\label{app:maj_three_spin}
We here compute the CMI of the three-spin distribution at effective temperature $x$,
\begin{equation}
\label{eq:app_three_spin_distribution}
\begin{gathered}
p_x(a,b,c)=\frac{1}{Z_x}
\exp\left[\frac{1}{x}(ab+bc+ca)\right],\\
Z_x=2\left(e^{3/x}+3e^{-1/x}\right).
\end{gathered}
\end{equation}
It is convenient to factor out $e^{3/x}$ and set $\kappa=e^{-4/x}$,
so that a configuration with all three spins aligned ($a = b = c$) has relative weight $1$,
while each of the six configurations with one spin opposite to the other two has
relative weight $\kappa$.
 
Since $p_x$ is invariant under flipping all three spins, the conditional
distribution of $(a,c)$ is the same for both values of $b$, and we may set
$b=+1$.  Of the four configurations of $(a,c)$, only $a=c=+1$ is fully
aligned.  
With $r={\kappa}/{(1+3\kappa)}$, 
the conditional distribution is therefore
\begin{equation}
\begin{array}{c|cc}
 & c=+1 & c=-1\\ \hline
a=+1 & 1-3r & r\\
a=-1 & r & r
\end{array}.
\end{equation}
Each side spin is opposite to $b$ with probability $2r$, so
$I(a:c|b)=H(a|b)+H(c|b)-H(a,c|b)$ gives
\begin{equation}
\label{eq:maj_scaling_functions}
\mathrm{CMI}(x;3,1)
=2h_2(2r)+(1-3r)\ln(1-3r)+3r\ln r ,
\end{equation}
with $h_2$ the binary entropy.  Substituting $r=\kappa/(1+3\kappa)$, so that
$1-3r=1/(1+3\kappa)$, gives Eq.~\eqref{Eq:three-spin-cmi}.

\subsection{MI and CMI for the local tripartition}
\label{app:maj_local_cmi}
We now derive Eq.~\eqref{eq:maj_local_cmi_scaling}, which corresponds to the CMI for which $N_A=N_C=1$ and
$N_B=N-2$.  
Consider one majority block at temperature $y$. Denote one child spin of interest by $a$
and the other two by $u$ and $v$. 
Conditioned on the parent spin $s$, the block
distribution is
\begin{equation}
P_y(a,u,v|s)=
\frac{\delta_{s,\operatorname{maj}(a,u,v)}
\exp[(au+uv+va)/y]}
{e^{3/y}+3e^{-1/y}} .
\end{equation}
If $u=v$, the parent majority spin is already fixed by $u$ and $v$, so $a$
cannot transmit CMI to the coarser problem.  If $u\ne v$, the parent spin equals
$a$, so $a$ survives as the side variable of the coarser problem.  
Therefore, the
probability that $a$ still determines the majority spin of its block is 
\begin{equation}
\label{eq:app_side_spin_probability}
\begin{aligned}
P_y(u\ne v|s)
&=\sum_{a,u,v}\mathbf{1}_{u\ne v}P_y(a,u,v|s)\\
&=\frac{e^{-1/y}+e^{-1/y}}{e^{3/y}+3e^{-1/y}}
=\frac{2e^{-4/y}}{1+3e^{-4/y}} .
\end{aligned}
\end{equation}
This must happen in both blocks containing $A$ and $C$ at the same RG step, so
one step contributes
\begin{equation}
\label{eq:app_side_spin_factor}
q(y)=P_y(u\ne v|s)^2
=\left(\frac{2e^{-4/y}}{1+3e^{-4/y}}\right)^2 .
\end{equation}
Iterating from the microscopic scale to the top RG scale gives
\begin{equation}
\mathrm{CMI}(T;N,N-2)
=\mathrm{CMI}(x;3,1)
\prod_{r=1}^{n-1}q\!\left(\frac{x}{3^{r/\nu}}\right),
\end{equation}
which is Eq.~\eqref{eq:maj_local_cmi_scaling}.

We now derive the asymptotic UV ($x\ll1$) and IR ($x\gg1$) limits. 
The first factor $\mathrm{CMI}(x;3,1)$ follows
from Eq.~\eqref{eq:maj_scaling_functions}, which scales as
$(4/x)e^{-4/x}$ for $x\ll1$ and as $1/(2x^2)$ for $x\gg1$.
On the other hand, when $x\ll1$, every term in $\prod_{r=1}^{n-1}q\!\left(\frac{x}{3^{r/\nu}}\right)$ is in the low-temperature regime $q(y)\sim 4e^{-8/y}$, $y\to0$, and thus
\begin{equation}
\prod_{r=1}^{n-1}q\!\left(\frac{x}{3^{r/\nu}}\right)
\sim (N/3)^{\log_3 4}
\exp\!\left[
-\frac{8}{T}\frac{1-(N/3)^{-1/\nu}}{1-3^{-1/\nu}}
\right].
\end{equation}
Together with $\mathrm{CMI}(x;3,1)\sim4e^{-4/x}/x$ and
$x=T(N/3)^{1/\nu}$, this gives
Eq.~\eqref{eq:maj_local_cmi_asymptotics}.
 
Now consider the IR regime $x\gg1$.  Reindexing the product by
$k=n-1-r$ gives 
\begin{equation}
\prod_{r=1}^{n-1}q\!\left(\frac{x}{3^{r/\nu}}\right)
=4^{-(n-1)}\prod_{k=0}^{n-2}4q\!\left(T3^{k/\nu}\right).
\end{equation}
Since $4q(y)=1-2/y+O(y^{-2})$ as $y\to\infty$, the infinite product $R_{\mathrm{CMI}}(T)=\prod_{k=0}^{\infty}4q\!\left(T3^{k/\nu}\right)$
converges, and therefore
\begin{equation}
\prod_{r=1}^{n-1}q\!\left(\frac{x}{3^{r/\nu}}\right)
\sim R_{\mathrm{CMI}}(T)(N/3)^{-\log_3 4}.
\end{equation}
Combining this with
$\mathrm{CMI}(x;3,1)\sim1/[2T^2(N/3)^{2/\nu}]$ gives the
IR result in
Eq.~\eqref{eq:maj_local_cmi_high_temperature}.

For comparison, we also study the mutual information between $A$ and $C$.
The two top blocks containing these spins have
correlation
\begin{equation}
\rho_x=\langle a c\rangle_x
=\frac{1-e^{-4/x}}{1+3e^{-4/x}},
\end{equation}
where the average is taken in the three-spin distribution in
Eq.~\eqref{eq:app_three_spin_distribution}.  A single microscopic spin $a$ has
conditional correlation with its parent
\begin{equation}
\begin{aligned}
\lambda(y)=\langle a s\rangle_{y,s}
&=P_y(a=s|s)-P_y(a=-s|s)\\
&=\frac{1+e^{-4/y}}{1+3e^{-4/y}} .
\end{aligned}
\end{equation}
Since the branches containing $A$ and $C$ are conditionally independent given the top spins, their correlation is
\begin{equation}
\rho_N(T)=\rho_x\prod_{r=1}^{n-1}
\lambda\!\left(\frac{x}{3^{r/\nu}}\right)^2 .
\end{equation}
Furthemore, because $A$ and $C$ are unbiased, their mutual information is
\begin{equation}
\begin{aligned}
\mathrm{MI}(T;N,N-2)
&=\frac{1+\rho_N}{2}\ln(1+\rho_N)\\
&\quad+\frac{1-\rho_N}{2}\ln(1-\rho_N).
\end{aligned}
\end{equation}
At low temperature,
\begin{equation}
1-\lambda(y)\sim 2e^{-4/y},\qquad y\to0,
\end{equation}
so the infinite product
\begin{equation}
\rho_\infty(x)=\rho_x\prod_{r=1}^{\infty}
\lambda\!\left(\frac{x}{3^{r/\nu}}\right)^2
\end{equation}
converges at fixed $x$.  Therefore $\rho_N\to\rho_\infty(x)$ as $N\to\infty$ at
fixed $x$, and $\mathrm{MI}(T;N,N-2)$ has a standard scaling collapse,
\begin{equation}
\begin{aligned}
\mathrm{MI}(T;N,N-2)\rightarrow
&\frac{1+\rho_\infty(x)}{2}\ln(1+\rho_\infty(x))\\
&+\frac{1-\rho_\infty(x)}{2}\ln(1-\rho_\infty(x)).
\end{aligned}
\end{equation}
This is in sharp contrast to the CMI: the ordinary correlation between $A$ and
$C$ passes through an ordered majority block with $\lambda(y)\simeq1$, whereas
their conditional correlation survives only when both side spins still
determine their block majorities, which happens with probability
$q(y)\sim4e^{-8/y}$.

\section{Derivation of Eq.~\eqref{eq:stable_tree_bmax}}
\label{app:stable_tree_bmax_counting}
 
We here derive Eq.~\eqref{eq:stable_tree_bmax} for the largest buffer that still leaves the zero-noise CMI equal to $\log 2$.  
It is more illuminating to first write the argument for a rule with repetition-block size $b$ and hence branching number
$2b$, where the stable tree in the main text has $b=3$. 

Let $N=(2b)^n$, and let $B_n$ be the largest buffer at depth $n$ for
which one parity constraint still links $A$ and $C$ at zero noise.  At depth
one there is no nonempty buffer with this property, so $B_1=0$.  When
we add one tree layer, the previous buffer appears inside the block at the
center.  We may also add $b$ complete child blocks on each side of the center.
Each such child block has size $(2b)^{n-2}$, so
\begin{equation}
B_n=B_{n-1}+2b(2b)^{n-2}.
\end{equation}
Iterating the recursion gives
\begin{equation}
B_n
=2b\sum_{r=0}^{n-2}(2b)^r
=\frac{(2b)^n-2b}{2b-1}
=\frac{N-2b}{2b-1}.
\end{equation}
Thus each new layer adds $b$ complete blocks on each side of the previous
buffer.  For the tree used in the main text, $b=3$ and hence
$B_{\text{max}}=(N-6)/5$.

\section{A nonideal majority--parity RG for the SSB--SWSSB tree}
\label{app:tree_nonideal_rg}

This appendix analyzes the simpler (but nonideal) RG briefly mentioned in
Sec.~\ref{sec:tree_ideal_RG}. 
Within each repetition triple, the map keeps the
majority sign and then multiplies the two resulting signs. 
This gives an exact
recursion for the independent noise strength, and repeated application removes
sufficiently weak noise.  However, we will show that this majority-parity RG discards correlations and is thus
not an ideal RG.

\subsection{RG flows for the SSB and SWSSB fixed points}

We first derive the noise recursions for the two elementary tree rules in
Sec.~\ref{sec:two_elementary_fixed}.  For the SSB repetition rule, consider
applying independent spin-flip noise of strength $p$.  Majority vote infers the
wrong parent only if exactly two or all three children are flipped.  These
events have probabilities $3p^2(1-p)$ and $p^3$, respectively, so
\begin{equation}
\label{Eq:p_maj}
p'_{\mathrm{maj}}
=p^3+3p^2(1-p)
=3p^2-2p^3
=O(p^2).
\end{equation}
The absence of a linear term means that majority vote suppresses weak noise
and flows toward the noiseless SSB state, as shown in
Fig.~\ref{fig:tree_nonideal_rg_flows}(a).

For the SWSSB parity rule, the parent is the product of its two children.  This
product changes sign only when exactly one child is flipped, and hence
\begin{equation}
\label{Eq:p_parity}
p'_{\mathrm{parity}}
=2p(1-p)
=2p-2p^2
=2p+O(p^2).
\end{equation}
The slope at $p=0$ is $2>1$, so weak noise grows and the flow approaches
$p=1/2$, as shown in Fig.~\ref{fig:tree_nonideal_rg_flows}(b). 

\subsection{RG flow for the SSB-SWSSB tree}

We now apply independent spin-flip noise of strength $p$ to the
SSB--SWSSB tree generated by Eq.~\eqref{Eq:stable_tree_rule}.  Since the
tree rule first applies parity and then repeats each logical child three times,
a natural coarse graining reverses these steps: it first takes a majority vote
within each triple and then multiplies the two majority outputs.

By Eq.~\eqref{Eq:p_maj}, either majority output is wrong with probability
$p_{\mathrm{maj}}=p^3+3p^2(1-p)$.  Their product is wrong exactly when one of
the two outputs is wrong.  The resulting six-to-one recursion is
\begin{equation}
\label{Eq:stable_tree_majority_recursion}
\begin{aligned}
p'&=2p_{\mathrm{maj}}(1-p_{\mathrm{maj}})\\
&=2\bigl(p^3+3p^2(1-p)\bigr)
\bigl(1-p^3-3p^2(1-p)\bigr).
\end{aligned}
\end{equation}
Let $\rho(p,N)$ denote the state on $N$ leaves after independent spin-flip
noise of strength $p$, and write $\rho(p)$ when the system size is unimportant.
If $\mathcal E$ denotes the majority--parity channel, this family is closed
exactly:
\begin{equation}
\mathcal E\!\left[\rho(p,N)\right]=\rho(p',N/6).
\end{equation}
Thus Eq.~\eqref{Eq:stable_tree_majority_recursion} is an exact recursion for the
noise strength.  Its beta function is
\begin{equation}
\beta(p)\equiv p'-p
=-p(1-2p)\left(1-4p^4+10p^3-4p^2-4p\right).
\end{equation}
Besides the stable fixed points $p=0$ and $p=1/2$, the map has one unstable
fixed point,
\begin{equation}
p_c^{\mathrm{nonideal}}\simeq0.225.
\end{equation}

\begin{figure*}[t]
\centering
\begin{minipage}[t]{0.31\textwidth}
\centering
\begin{tikzpicture}[x=8.2cm,y=1cm,>=Stealth]
\draw[->] (-0.02,0) -- (0.55,0) node[right] {$p$};
\draw (0,0.07) -- (0,-0.07) node[below] {$0$};
\draw (0.5,0.07) -- (0.5,-0.07) node[below] {$1/2$};
\draw[very thick,blue,->] (0.43,0.18) -- (0.30,0.18);
\draw[very thick,blue,->] (0.30,0.18) -- (0.17,0.18);
\draw[very thick,blue,->] (0.17,0.18) -- (0.04,0.18);
\filldraw[black] (0,0) circle (1.8pt);
\draw[black,fill=white] (0.5,0) circle (1.8pt);
\node[above] at (0.25,0.42) {$\beta(p)<0$};
\node at (0.25,-0.68) {(a) SSB};
\end{tikzpicture}
\end{minipage}
\hfill
\begin{minipage}[t]{0.31\textwidth}
\centering
\begin{tikzpicture}[x=8.2cm,y=1cm,>=Stealth]
\draw[->] (-0.02,0) -- (0.55,0) node[right] {$p$};
\draw (0,0.07) -- (0,-0.07) node[below] {$0$};
\draw (0.5,0.07) -- (0.5,-0.07) node[below] {$1/2$};
\draw[very thick,red,->] (0.05,0.18) -- (0.18,0.18);
\draw[very thick,red,->] (0.18,0.18) -- (0.31,0.18);
\draw[very thick,red,->] (0.31,0.18) -- (0.46,0.18);
\draw[black,fill=white] (0,0) circle (1.8pt);
\filldraw[black] (0.5,0) circle (1.8pt);
\node[above] at (0.25,0.42) {$\beta(p)>0$};
\node at (0.25,-0.68) {(b) SWSSB};
\end{tikzpicture}
\end{minipage}
\hfill
\begin{minipage}[t]{0.31\textwidth}
\centering
\begin{tikzpicture}[x=8.2cm,y=1cm,>=Stealth]
\draw[->] (-0.02,0) -- (0.55,0) node[right] {$p$};
\draw (0,0.07) -- (0,-0.07) node[below] {$0$};
\draw (0.225,0.07) -- (0.225,-0.07);
\node at (0.225,-0.24) {$p_c$};
\draw (0.5,0.07) -- (0.5,-0.07) node[below] {$1/2$};
\draw[very thick,blue,->] (0.19,0.18) -- (0.11,0.18);
\draw[very thick,blue,->] (0.11,0.18) -- (0.035,0.18);
\draw[very thick,red,->] (0.265,0.18) -- (0.36,0.18);
\draw[very thick,red,->] (0.36,0.18) -- (0.47,0.18);
\filldraw[black] (0,0) circle (1.8pt);
\draw[black,fill=white] (0.225,0) circle (1.8pt);
\filldraw[black] (0.5,0) circle (1.8pt);
\node[above] at (0.11,0.42) {$\beta(p)<0$};
\node[above] at (0.37,0.42) {$\beta(p)>0$};
\node at (0.25,-0.68) {(c) SSB--SWSSB tree};
\end{tikzpicture}
\end{minipage}
\caption{RG flows for the SSB, SWSSB, and SSB-SWSSN tree fixed points.
(a) Majority vote for the SSB repetition rule flows to $p=0$.  (b) The product
map for the SWSSB parity rule flows to $p=1/2$.  (c) The combined
majority--parity map flows to $p=0$ below
$p_c=p_c^{\mathrm{nonideal}}$ and to $p=1/2$ above it.}
\label{fig:tree_nonideal_rg_flows}
\end{figure*}
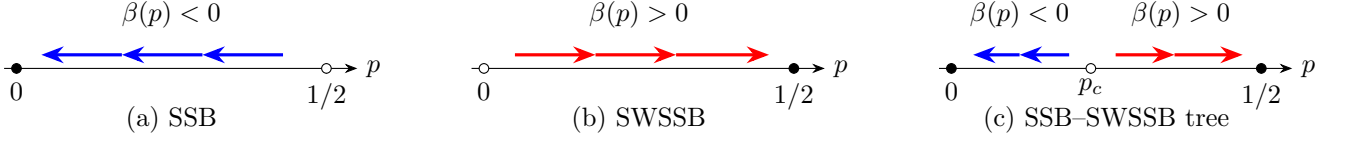

For small $p$, majority vote first reduces the error to
$p_{\mathrm{maj}}=O(p^2)$, after which the parity step gives
$p'=6p^2+O(p^3)$.  Thus $p=0$ remains stable.  The intermediate fixed point
$p_c^{\mathrm{nonideal}}$ separates the basin flowing to $p=0$ from the basin
flowing to $p=1/2$, as summarized in
Fig.~\ref{fig:tree_nonideal_rg_flows}(c).

\subsection{Majority--parity RG is nonideal}

Although majority vote gives an exact recursion for $p$, it does not preserve
all correlations with the rest of the tree.  It retains the majority sign but
forgets whether the vote was unanimous or split two to one.  To see why this
matters, let $L=(s_1,s_2,s_3)$ and $R=(s_4,s_5,s_6)$ be the two noisy triples
in one block, and condition on the parent spin $s=+1$.
Equation~\eqref{Eq:stable_tree_rule} then gives $s_L=s_R$, with their common
value equally likely to be $+1$ or $-1$.
 
If keeping only $\operatorname{maj}(L)$ preserved the correlations between the
two triples, then 
\begin{equation}
I(L:R|s=+1)=I(\operatorname{maj}(L):R|s=+1).
\end{equation}
This equality fails.  For example, $L=(+1,+1,+1)$ and
$L=(+1,+1,-1)$ have the same majority but give different posterior
probabilities for the logical spin $s_L$:
\begin{equation}
\begin{aligned}
P(s_L=+1|L=(+1,+1,+1))
&=\frac{(1-p)^3}{(1-p)^3+p^3},\\
P(s_L=+1|L=(+1,+1,-1))
&=1-p.
\end{aligned}
\end{equation}
For $0<p<1/2$, the unanimous observation is stronger evidence for
$s_L=+1$.  Since the noiseless spins satisfy $s_R=s_L$, the two observations
also give different predictions for $R$.  Majority vote maps both to $+1$ and
erases this difference.  The subsequent parity step cannot restore the
discarded information, so the combined majority--parity map is nonideal.

\subsection{Usefulness of nonideal RG}
  
Although the RG is nonideal, it still provides useful information about mixed-state phases based on two-way connections~\cite{sang2024mixed}.
In particular, for $p<p_c^{\mathrm{nonideal}}$, repeated application of the RG gives a local-channel
path $\rho(p)\to\rho(0)$. Independent spin-flip noise gives a local-channel
path in the opposite direction, $\rho(0)\to\rho(p)$. 
Therefore, the nonideal RG shows that any state with $p<p_c^{\mathrm{nonideal}}$ is connected to $\rho(0)$ in both directions through local-channel transformations.

\section{Exact recursion for the ideal tree RG}
\label{app:ideal_tree_rg}
 
We now derive the exact recursion used in the ideal RG. Let $S$ denote the
observations in a block and $\sigma$ its hidden parent spin. We summarize what
$S$ tells us about $\sigma$ by the posterior probability
\begin{equation}
q(S)=P(\sigma=+1|S)
=\frac{P(S|\sigma=+1)}{P(S|\sigma=+1)+P(S|\sigma=-1)},
\end{equation}
where we have used the equal prior on $\sigma=\pm1$. Thus a single number
$q(S)\in[0,1]$ describes what the observations tell us about $\sigma$. In particular, at a
noisy leaf,
\begin{equation}
q^{(0)}(+1)=1-p,\qquad q^{(0)}(-1)=p.
\end{equation}

We next determine the posterior probability for a parent block from those of
its six child blocks. Let
$q_j=P(\sigma_j=+1|s_j)$ be the posterior probability for child block $j$. 
For the left repetition triple, let
$a$ be its logical spin. The repetition constraint sets
$\sigma_1=\sigma_2=\sigma_3=a$.
Since the child blocks are conditionally independent given $a$, their
posterior odds multiply:
\begin{equation}
\frac{P(a=+1|s_1,s_2,s_3)}{P(a=-1|s_1,s_2,s_3)}
=\prod_{j=1}^3\frac{q_j}{1-q_j}.
\end{equation}
Converting these odds back to a probability gives
\begin{equation} 
M(q_1,q_2,q_3)
:=
\frac{q_1q_2q_3}
{q_1q_2q_3+(1-q_1)(1-q_2)(1-q_3)}.
\end{equation}
For a full six-child block, let $a$ and $b$ be the left and right repetition
logical spins.  Their posteriors are
\begin{equation}
\begin{gathered}
\alpha=M(q_1,q_2,q_3),\qquad
\beta=M(q_4,q_5,q_6).
\end{gathered}
\end{equation}
With independent uniform priors for $a$ and $b$, their posterior factorizes
after conditioning on the two triples.  The parent spin is $\sigma=ab$, so
$\sigma=+1$ exactly when $a=b$.  Therefore
\begin{equation}
\label{Eq:stable_tree_ideal_channel}
\mathcal E_{\mathrm{ideal}}(q_1,\ldots,q_6)
=:G(q_1,\ldots,q_6)
=\alpha\beta+(1-\alpha)(1-\beta).
\end{equation}
Applying this rule at every level gives the exact recursion
\begin{equation}
\label{Eq:stable_tree_posterior_recursion}
q^{(h+1)}(s_1,\ldots,s_6)
=\mathcal E_{\mathrm{ideal}}\!\left(
q^{(h)}(s_1),\ldots,q^{(h)}(s_6)\right),
\end{equation}
where each $s_i$ denotes the observations in one depth-$h$ child block.

\end{document}